\documentclass[usenatbib, usegraphicx, useAMS]{mn2e}

\usepackage{amsmath}
\usepackage{color}
\usepackage{graphicx}
\usepackage{hyperref}

\newcommand{\peakCl}{{C_{\ell}^{\text{peak}}}}   
\newcommand{\Npix}{\ensuremath{{N_{\text{pix}}}}}
\newcommand{\Nside}{\ensuremath{{N_{\text{side}}}}}
\newcommand{\lmax}{\ensuremath{{\ell_{\text{max}}}}}
\newcommand{\lmin}{\ensuremath{{\ell_{\text{min}}}}}
\newcommand{\fsky}{\ensuremath{{f_{\text{sky}}}}}
\newcommand{\Normal}{\ensuremath{{\cal{N}}}}
\newcommand{\perplex}{\ensuremath{\mathcal{P}}}

\title[TEASING: fast low-$\ell$ likelihood for CMB
temperature]{TEASING: a fast and accurate approximation for the low
  multipole likelihood of the Cosmic Microwave Background temperature}

\author[K.~Benabed, J.-F.~Cardoso, S.~Prunet \& E.~Hivon]{
K.~Benabed$^1$\thanks{E-mail: \href{mailto:benabed@iap.fr}{{benabed@iap.fr}}},
J.-F.~Cardoso$^{1,2}$,
S.~Prunet$^1$ \& 
E.~Hivon$^1$ \\
$^1$Institut d'Astrophysique de Paris, 98bis Bd Arago, 75014 Paris, France.\\
$^2$Laboratoire de Traitement et Communication de l'Information, LTCI/CNRS 46, rue Barrault, 75013 Paris, France.}
\date{Accepted ---. Received ---; in original form \today}
\pagerange{\pageref{firstpage}--\pageref{lastpage}}
\pubyear{2009}

\begin{document}

\label{firstpage}
\maketitle

\begin{abstract}
  We explore the low-$\ell$ likelihood of the angular spectrum
  $C_\ell$ of masked CMB temperature maps using an adaptive importance
  sampler.  We find that, in spite of a partial sky coverage, the
  likelihood distribution of each $C_{\ell}$ closely follows
  an inverse gamma distribution.  Our exploration is accurate enough
  to measure the inverse gamma parameters along with the correlation
  between multipoles.  Those quantities are used to build an
  approximation of the joint posterior distribution of the low-$\ell$
  likelihood.  The accuracy of the proposed approximation is
  established using both statistical criteria and a mock cosmological
  parameter fit.  When applied to the WMAP5 data set, this
  approximation yields cosmological parameter estimates at the
  same level of accuracy as the best current techniques but with very
  significant speed gains.
\end{abstract}


\section{Introduction}

The CMB angular spectrum $C=\{C_\ell\}$ is a central quantity for
conducting statistical inference based on CMB observations~\citep{Bond:1987MNRAS.226..655B}.
The high resolution of available \citep{Hinshaw:2008p6414} and
forthcoming CMB observations \citep{bluebook} makes it necessary (at
least in the case of partial sky coverage) to adopt a processing
scheme in which the low-$\ell$ and high-$\ell$ parts of the data are
processed independently \citep{Efstathiou:2006p4428}.
This paper addresses the large scale part of the problem: inference
regarding low multipoles based on a partial low-resolution CMB map.

After defining the problem of low-$\ell$ pixel-based likelihood and
introducing some notations (Sec.~\ref{sec:likelihood}), we first show
how to build a (large) set of $N$ \emph{importance samples} of the angular
spectrum such that all integrals of interest for statistical inference
can be approximated by Monte-Carlo estimates
(Sec.~\ref{sec:building-samples}).
Based on those results, we propose in Sec.~\ref{sec:lapprox} a new
approximation to the likelihood for partially observed low-resolution
CMB maps.  This approximation was initially built as part of the
importance sampler but it turns out to be so accurate that it is of
independent interest.
This paper and the recent reference~\citep{Rudjord:2008p4864} are
similar in spirit but differ in the sampling method and in the
proposed likelihood approximation.

\section{Likelihood}\label{sec:likelihood}

We recall some well-known facts about the likelihood of the angular
spectrum of a CMB temperature map.

In the ideal case of noise-free, beam-free, full-sky map
(represented by the vector $\mathbf{x}$ of pixels), one has
direct access to the harmonic coefficients $a_{\ell m}$ of the sky.
Assuming an isotropic Gaussian field, the empirical angular spectrum
$\widehat{C}_{\ell}=\frac{1}{2\ell+1}\sum_{m}|a_{\ell m}|^{2}$ is a
sufficient statistic for the data and their probability distribution
takes the factorized form~\citep{BJK:2000ApJ...533...19B}:
\begin{equation}
  \label{eq:likely-factorized}
  p(\mathbf{x}|C)
  \propto
  \prod_{\ell\geq0}\exp-\frac{2\ell+1}{2}
  \left(\frac{\hat C_\ell }{C_{\ell}}+\log C_{\ell}\right)
  .
\end{equation}
In the case of a flat prior $p(C)$,
expression~(\ref{eq:likely-factorized}) combined with Bayes rule
$p(C|\mathbf{x})=p(\mathbf{x}|C)p(C)/p(\mathbf{x})$ reveals that,
given $\mathbf{x}$, the angular spectrum $C$ is distributed as a
product of inverse gamma densities:
\begin{align}
  \label{eq:factorpost}
  p(C|\mathbf{x})         & =  \prod_{\ell}i\Gamma(C_{\ell};\alpha_{\ell},\beta_{\ell})\\
  i\Gamma(x;\alpha,\beta) & \equiv  \frac{\beta^{\alpha}}{\Gamma(\alpha)}\ x^{-\alpha-1}\   e^{-\frac{\beta}{x}},
  \label{eq:defigamma}
\end{align} 
with parameters $\alpha_{\ell}=(2\ell-1)/2$ and
$\beta_{\ell}=(2\ell+1)\hat C_\ell/2$.

Such a factorization does not hold when only a fraction of the sky is
observed (or has to be ignored because of excessive contamination by
foregrounds), or when the stationary CMB is contaminated by non
stationary
noise~\citep{Gorski:1994ApJ...430L..85G,Tegmark:1997PhRvD..55.5895T}.
However, for small sky masks and/or small deviations from
stationarity, deviations from the factorized form
(\ref{eq:likely-factorized}) are expected to be small, suggesting the
new likelihood approximation developed in Sec.~\ref{sec:lapprox}.

\bigskip
\par\noindent\textbf{Pixel-based likelihood.}
We turn to the actual case of interest: partial sky coverage, presence
of independent additive Gaussian noise, low-pass effect of a beam.
The data set, represented by an $\Npix\times 1$ vector $\mathbf{x}$ of
pixel values, can no longer be losslessly compressed into a sufficient
spectral statistic $\hat C_\ell$.  Rather, one must use the plain
Gaussian density:
\begin{equation}
  p(\mathbf{x}| \mathbf{R})
  =
  |2\pi\mathbf{R}|^{-1/2}e^{-\frac{1}{2}\mathbf{x}^{T}\mathbf{R}^{-1}\mathbf{x}}
  \label{eq:likely-general}
\end{equation}
where the covariance matrix $\mathbf{R}$ of $\mathbf{x}$ has
contributions from the CMB signal and from noise.  
For two pixels $i$ and $j$ with angular separation $\theta_{ij}$, the
CMB part of the covariance matrix has an $(i,j)$ entry given by~\citep{BJK:2000ApJ...533...19B}
\begin{equation}  
  \sum_\ell \frac{2\ell+1}{4\pi}W_{\ell}C_{\ell}P_{\ell}(\cos\theta_{ij})
\end{equation}
where $P_{\ell}$ is the Legendre polynomial of order~$\ell$ and where
the window function $W_{\ell}$ can represent \textit{e.g.} the
spectral response of an azimuthally symmetric beam, or more generally
the convolution of the signal with any azimuthally symmetric
kernel. Hence, we ignore the complications due to an anisotropic beam
as well as the presence of residual foreground contaminants.

The noise part of the covariance matrix could take any form but, in
this work, it is taken to correspond to an isotropic noise with
angular spectrum $N_\ell$.  We can thus define a total angular
spectrum $D_{\ell}$
\begin{equation}\label{eq:CtoD}
  D_{\ell} = W_{\ell}C_{\ell}+N_{\ell} 
\end{equation} 
which is unambiguously related to $C_\ell$ since the beam $B_\ell$ and
the noise spectrum $N_\ell$ are assumed to be known.

\bigskip
\par\noindent\textbf{Free parameters.}
In practice, we consider a more restricted model for the covariance
matrix of the observed pixels.  
First, the adjustable multipoles are restricted to a range
$\lmin\leq\ell\leq\lmax$ while other multipoles are kept at constant
values.
Second, we only consider uncorrelated noise with zero mean and
variance $\sigma^2$ per pixel.  It contributes a term
$\sigma^2\delta_{ij}$ to $\mathbf{R}$ and corresponds to a flat
angular spectrum $N_\ell = \sigma^2\,/ \Omega_\mathrm{pix}$ if all
pixels have the same area $\Omega_\mathrm{pix}$.
Then, the covariance matrix of $\mathbf{x}$ as a function of
$\mathbf{D}=\{D_\ell\}_{\ell=\lmin}^{\ell=\lmax}$ is spelled out as:
$\mathbf{R}(\mathbf{D}) = \mathbf{R}^\text{var} (\mathbf{D}) + \mathbf{R}^\text{cst}$ with
\begin{align}  
  \label{eq:covmatD}
  \mathbf{R}_{ij}^\text{var}(\mathbf{D}) &= 
  \sum_{\ell=\lmin}^{\ell=\lmax}  \frac{2\ell+1}{4\pi}(D_\ell-N_\ell)P_{\ell}(\cos\theta_{ij})  \\
  \mathbf{R}_{ij}^\text{cst} &= 
  \sum_{\ell\ \mathrm{fixed}} 
  \frac{2\ell+1}{4\pi} W_\ell C_\ell P_{\ell}(\cos\theta_{ij})  \ +\   \sigma^2 \delta_{ij}
  \label{eq:covmatDfix}
\end{align}

\bigskip
\par
\begin{figure}
  \centering
  \includegraphics[clip,width=1\columnwidth]{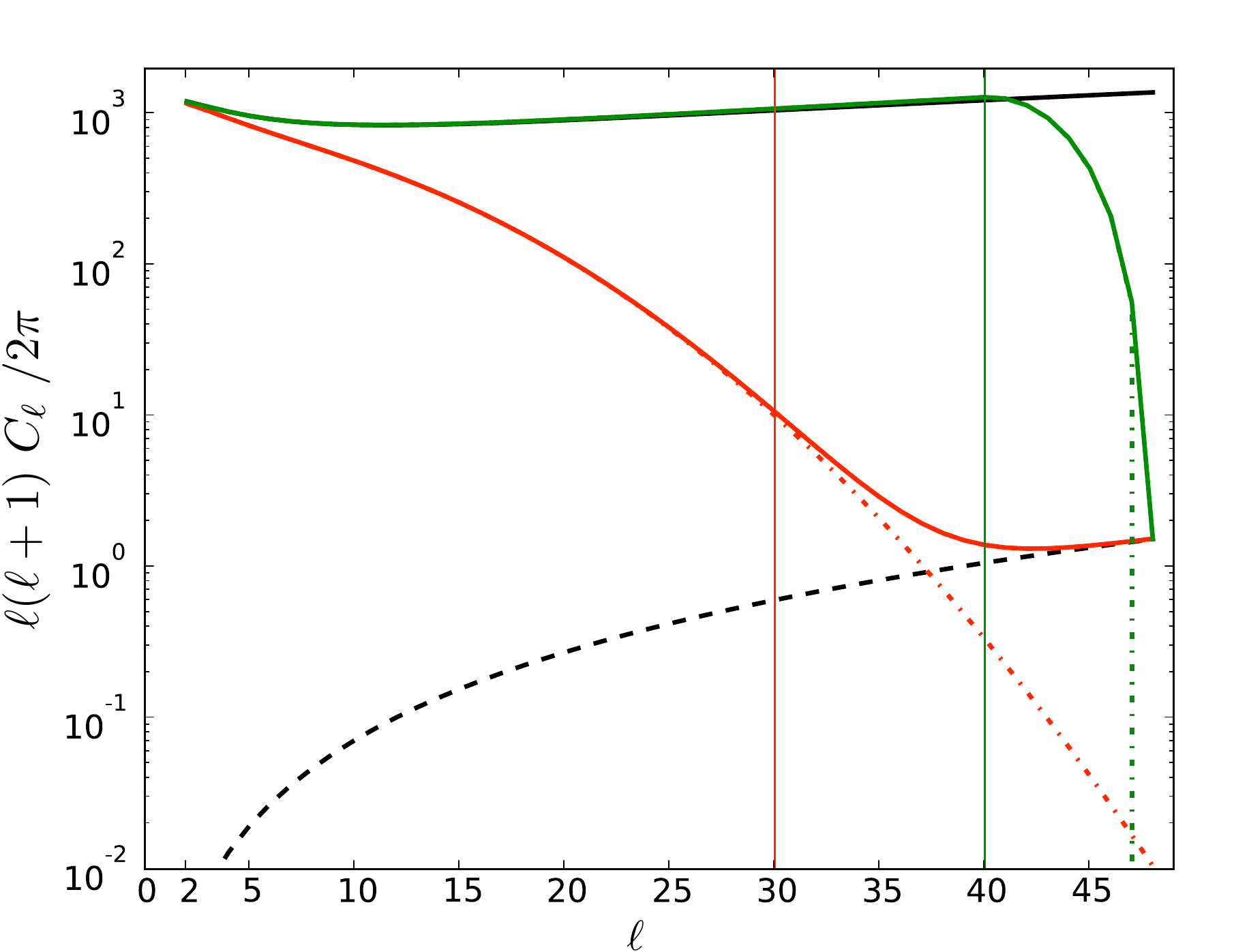}  
  \caption{The WMAP best fit spectrum $C_\ell$ (black solid line), the
    noise spectrum $N_\ell$ for a variance of $\sigma^2=1\mu
    K^2$/pixel (black dashed line), and the angular spectra $W_\ell
    C_\ell$ (dot dashed) and $W_\ell C_\ell + N_\ell$ (solid) when
    $W_\ell$ is the window function $W_\ell$ of eq.~(\ref{eq:archcos})
    (green) or the WMAP Gaussian beam (red).  Spectra are rescaled by
    $\ell(\ell+1)/2\pi$ for clarity.  }
  \label{fig:wlnlcl}
\end{figure}

\noindent\textbf{Priors and posterior distributions.}
In all the following, the prior distribution on $\mathbf{D}$ is taken to be
flat for $D_\ell\geq N_\ell$.  At all angular frequencies such that
$W_\ell C_\ell \gg N_\ell$ (figure~\ref{fig:wlnlcl} illustrates the
values used in this paper), this is almost identical to a flat prior
on the positive values of $C_\ell$.
The posterior distribution of $\mathbf{D}$ given the data $\mathbf{x}$ is
\begin{displaymath}
  \pi(\mathbf{D})= 
  p(\mathbf{D}|\mathbf{x}) 
  \propto
  p(\mathbf{x}| \mathbf{R}(\mathbf{D})) \ \prod_{\ell=\lmin}^{\ell=\lmax}  \mathbf{1}(D_\ell \geq N_\ell)
\end{displaymath}
where $p(\mathbf{x}| \mathbf{R}(\mathbf{D}))$ is evaluated using
eqs~(\ref{eq:likely-general}), (\ref{eq:covmatD}) and~(\ref{eq:covmatDfix}).

\bigskip\par\noindent\textbf{About noise and regularization.} 
On a cut sky, the CMB part of the covariance matrix may be poorly
conditioned with a trough in its eigenvalue spectrum corresponding to
those modes which are mostly localized in the cut.
In this case, it is
customary~\citep{Eriksen:2006p3364,Hinshaw:2007ApJS..170..288H} to add
a very small amount of noise to the data and to add the corresponding
contribution to the covariance matrix as in eq.~(\ref{eq:covmatDfix}).
Another reason for adding uncorrelated noise is to cover
spurious noise correlation possibly introduced when the observed sky
map is downgraded and to simplify the noise
structure~\citep{Dunkley:2008p3305}.
See figure~\ref{fig:wlnlcl} for the values used in our experiments.
Another possibility is regularization by projection onto the most
significant eigen-vectors of the covariance matrix
\citep{BJK:2000ApJ...533...19B} but this possibility is not considered
here.

\section{Building a sample of the low-$\ell$ posterior with importance sampling}
\label{sec:building-samples}

This section reports on the construction of \emph{importance samples} of
the $C_{\ell}$ under their joint posterior for two data sets.  The
principle of importance sampling is first briefly recalled in
section~\ref{sec:ISintro}; our specific technique (an adaptive
variant) is described in section~\ref{sub:adapt:algo} and applied to a
synthetic CMB cut sky map (sec.~\ref{sec:synthmap}) and to the
official WMAP5 low resolution map(sec.~\ref{sec:samplewmap5}).

\subsection{Importance sampling}\label{sec:ISintro}

Importance sampling is a well established technique to explore a
probability distribution when no method for directly sampling from it
is available (the well known VEGAS algorithm~\citep{Lepage:1978p4691}
for instance, is based on importance sampling).
Consider estimating the expectation $E f(x)=\int f(x)\pi(x)dx$ of some
function $f$ of $x$ when the random variable $x$ is distributed under
$\pi$.  If $x_i, i=1,N$ are $N$ samples of $x$, then $E f(x)$ can be
estimated by the sample average $\frac1N \sum_i f(x_i)$.  In contrast,
importance sampling relies on samples $x_i$ distributed under a
\emph{proposal distribution} $g$ not necessarily equal to $\pi$.  If the
support of $g$ includes the support of $\pi$ then
\begin{displaymath}
  E f = \int f(x) \pi(x) dx = \int f(x)\frac{\pi(x)}{g(x)} g(x) dx
\end{displaymath}
so that, if the samples $x_i$ are distributed under $g$, then $Ef$ is
estimated without bias by
\begin{displaymath}
  \frac1N \sum_{i=1}^N   w_i f(x_i)
  \quad\text{where}\quad
  w_i = w(x_i) \equiv \frac{\pi(x_i)}{g(x_i)}
\end{displaymath}
The factors $w_i$ are called \emph{importance weights}.

Monte-Carlo integration reaches its maximum efficiency when the
samples are drawn independently under a proposal distribution $g$
which is identical to the target distribution $\pi$.  While MCMC
methods try to draw from the target distribution $\pi$, they do not
build independent samples; in contrast, importance sampling (usually)
relies on independent draws from an approximate distribution $g$ and
corrects the discrepancy using importance weights $w_i$.  Therefore,
importance sampling should outperform MCMC methods whenever
independent samples can be drawn from a proposal distribution which is
``close enough'' to the target.

The agreement between target and proposal distributions can be
measured by the Kullback-Leibler divergence
\begin{equation}
  K(\pi|g)\equiv \int \log\frac{\pi(x)}{g(x)} \, \pi(x)\, dx ,
\end{equation} 
which is often remapped as the so-called \emph{perplexity criterion}:
$ \perplex(\pi|g)\equiv \exp - K(\pi|g) $ so that perfect agreement is
reached when $\perplex=1$.
Another criterion is the \emph{effective sample size} ($ESS$) of an
importance sample:
\begin{equation}
  ESS=\frac
  {\left(\sum_i w_i\right)^2 }
  {\sum_{i} w_i^2 }
\end{equation} 
If the proposal matches the target perfectly, then $ESS=N$, otherwise
it is smaller than the number of importance samples.  The effective
sample size is directly related to the variance of the MC estimates.

\bigskip

Importance sampling is well fitted to the problem at hand for at least
two reasons: ease of parallelization and availability of a good
proposal distribution.

Parallelization is a strong requirement due to the
high computational cost of CMB studies.
We are planning to sample a 30- to 40-dimensional space, and the
computation of the likelihood for a given angular spectrum costs about
5 seconds for $\lmax=48$ and $\Npix=3072$  on a typical 2GHz CPU. 
Since importance sampling can be trivially parallelized, it makes it
straightforward to take full advantage of CPU clusters.  For instance,
computing $10^5$ samples would take about 4 days on a single CPU but
is reduced to mere hours on a cluster.
The Markov-Chain Monte-Carlo algorithm cannot be parallelized as easily.
Indeed, to be able to mix different parallel chains, one has to ensure that 
they have correctly converged \citep{rosenthal00}, which can be a difficult task in 30 
to 40~dimensions.

Regarding the proposal distribution, one can draw inspiration from the
noise-free, full-sky case~(\ref{eq:factorpost}) since a mask hiding
less than $20\%$ of the sky and a high signal to noise situation are
expected to modify it only slightly\footnote{This situation is 
representative of CMB data sets from satellites such as WMAP and Planck.}.
Indeed, as demonstrated below, a product of independent inverse gamma
distributions turn out to be a very efficient proposal distribution,
provided it is correctly tuned.  Such a tuning is achieved via an
\emph{adaptive importance sampling}, as explained next.

\subsection{An adaptive importance sampling algorithm}\label{sub:adapt:algo}

Importance sampling is efficient only if the proposal distribution is
close enough to the target, an objective which may be difficult to
reach in large dimensions (sampling angular spectra in the range
$0\leq\ell\leq 40$ qualifies as large problem).  To tackle this
complexity, we resort to \emph{adaptive importance sampling} which consists
in running a sequence of importance runs in which the proposal
distribution is improved at each run based on the results of previous
runs.  
A more detailed description of adaptive importance sampling (based upon
the PMC algorithm from \cite{Cappe:2007p3640}) in the context of cosmology 
can be found in \cite{darren2009}.

\textbf{General scheme.}
The general scheme, based on a parametric family of proposal
distributions $g(\mathbf{y} ;\theta)$, is as follows:
\begin{enumerate}
\item\label{algo:init} Start with the best available guess of
  $\theta$ for the parameters of the proposal distribution.
\item\label{algo:debut} Sample under $g(\mathbf{y}; \theta)$. Compute
  and store the importance weights.
\item\label{algo:reest} Re-estimate $\theta$ so that $g(\mathbf{y};
  \theta)$ best matches the current sample set.
\item If the (estimated) perplexity $\perplex(\pi(\mathbf{y}) | g(\mathbf{y};
  \theta))$ is high enough (\textit{e.g.}  above $0.5$) or if it has
  not changed significantly during the last iterations, exit to
  \ref{algo:fin}.  Otherwise, go to \ref{algo:debut} for another
  importance run with the re-estimated parameters.
\item\label{algo:fin} Use the last value of $\theta$ for a large final
  importance sampling run.
\end{enumerate}

\textbf{Sampling angular spectra.}
In our experiments, we sample the total angular spectrum, that is,
${\mathbf{y}=\mathbf{D}} = \{ D_\ell\}_{\ell=\ell_\mathrm{min}}^{\ell=\ell_\mathrm{max}}$
and use independent inverse gamma distributions for the proposal:
\begin{displaymath}
  g( \mathbf{D} ; \theta ) = \prod_{\ell=\ell_\mathrm{min}}^{\ell=\ell_\mathrm{max}} 
  i\Gamma( D_\ell ;\alpha_\ell,\beta_\ell) .
\end{displaymath}
Hence we must adapt a vector $\theta=\{\alpha_\ell,
\beta_\ell\}_{\ell=\ell_\mathrm{min}}^{\ell=\ell_\mathrm{max}}$ of
$2(\ell_\mathrm{max}-\ell_\mathrm{min}+1)$ parameters.
As a starting point at step~\ref{algo:init}, we use
\begin{displaymath}
  \alpha_\ell =\frac{(2\ell+1)}{2}\fsky-1,  \qquad  
  \beta_\ell =\frac{(2\ell+1)}{2}\fsky  D_\ell^{ML}
\end{displaymath}
where $D_{\ell}^{ML}$ is the maximum likelihood estimate of the angular
spectrum.  At step~\ref{algo:reest}, parameters $\alpha_\ell$ and
$\beta_\ell$ are re-estimated at their maximum likelihood values (see appendix).

The target density $\pi(\mathbf{D})$ is the posterior distribution of
$\mathbf{D}$ when the prior distribution of $\mathbf{D}$ is flat.
Hence, it is proportional to the likelihood.

In the two examples presented below, this iterative algorithm reached
a perplexity above $0.6$ after the first step of 50k samples and a
500k samples set was produced during the final sampling phase.

\subsection{Synthetic map}\label{sec:synthmap}

\begin{figure}
  \includegraphics[width=1\columnwidth]{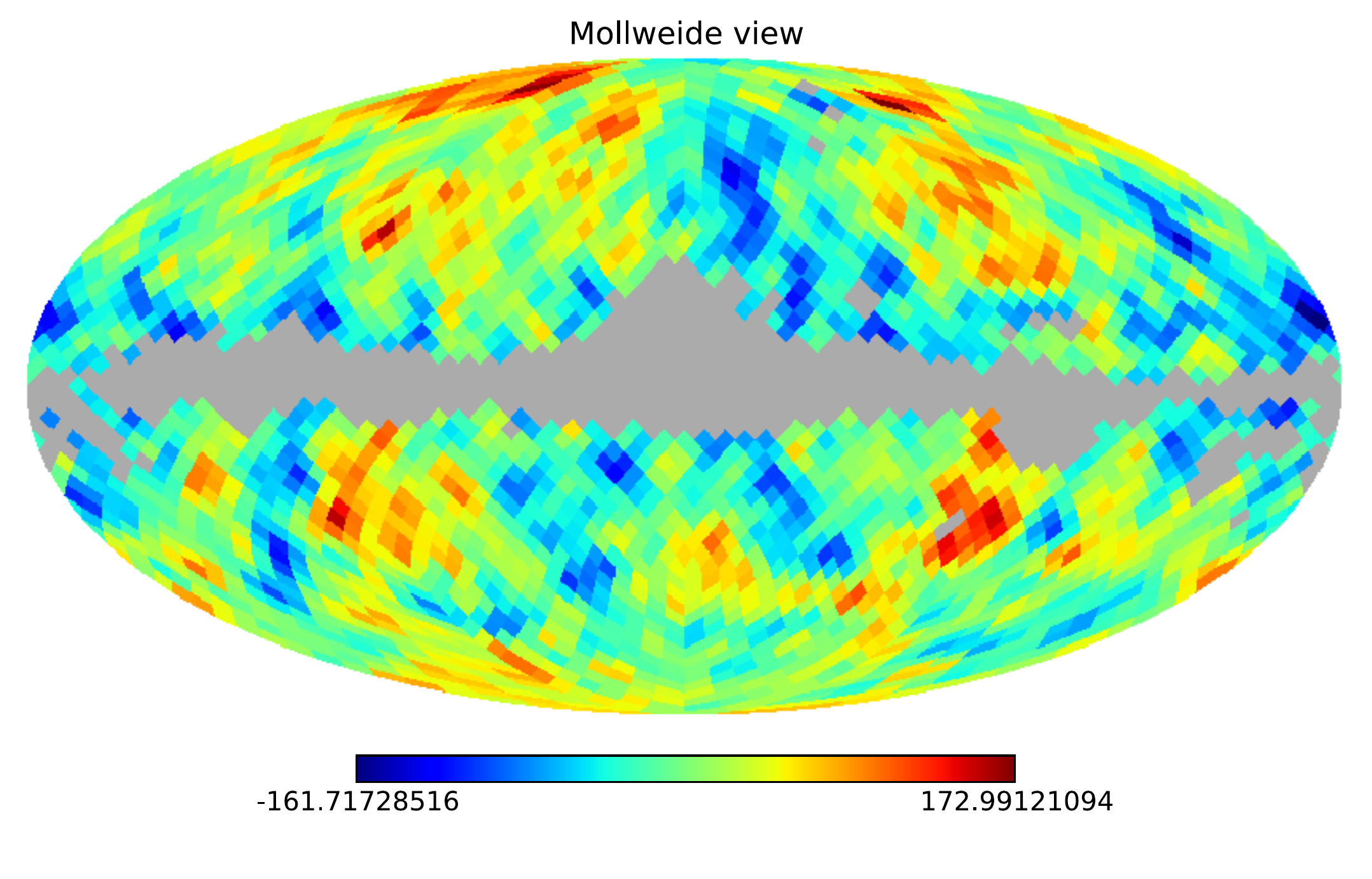}
  \caption{The synthetic CMB map used at sec.~\ref{sec:synthmap}.}
  \label{fig:syntmap}
\end{figure}

We first describe the results of adaptive importance sampling runs on
a synthetic CMB map.  The map is prepared at resolution $\Nside=16$
from the WMAP5 best fit power spectrum \cite{Dunkley:2008p3305} using
HEALPix~\citep{2005ApJ...622..759G}.  To avoid aliasing small
scale power into large scale modes, the map is smoothed prior
  to down-sampling using a synthetic window function $w_{\ell}$:
\begin{equation}\label{eq:archcos}
  W_{\ell}=\left\{ 
    \begin{array}{lc}
      1 & 0\leq \ell \leq 40\\
      \frac{1+\cos\left((\ell-40)\pi/8\right)}{2}      & 40\leq \ell\leq 48 \\
      0 & 48 \leq \ell 
    \end{array}\right.
\end{equation}
which is used to explore the posterior of $C_{\ell}$ up to $\ell=40$.
                                   
The posterior of the power spectrum is given by the likelihood
described in Eq.~(\ref{eq:likely-general}), with a flat prior.  The
Galactic region is excluded using the WMAP5 mask, hiding $18\%$ of the
sky.  The map is shown in figure~\ref{fig:syntmap}.  A $1\mu$K/pixel
noise is taken into account in the likelihood, but no noise is
actually added to the map.  This level should not affect our results
as it is much lower than $\Omega_\mathrm{pix} C_{40}$ (see figure \ref{fig:wlnlcl}).
\begin{figure*}
  \begin{centering}
    \includegraphics[clip,width=1\columnwidth]{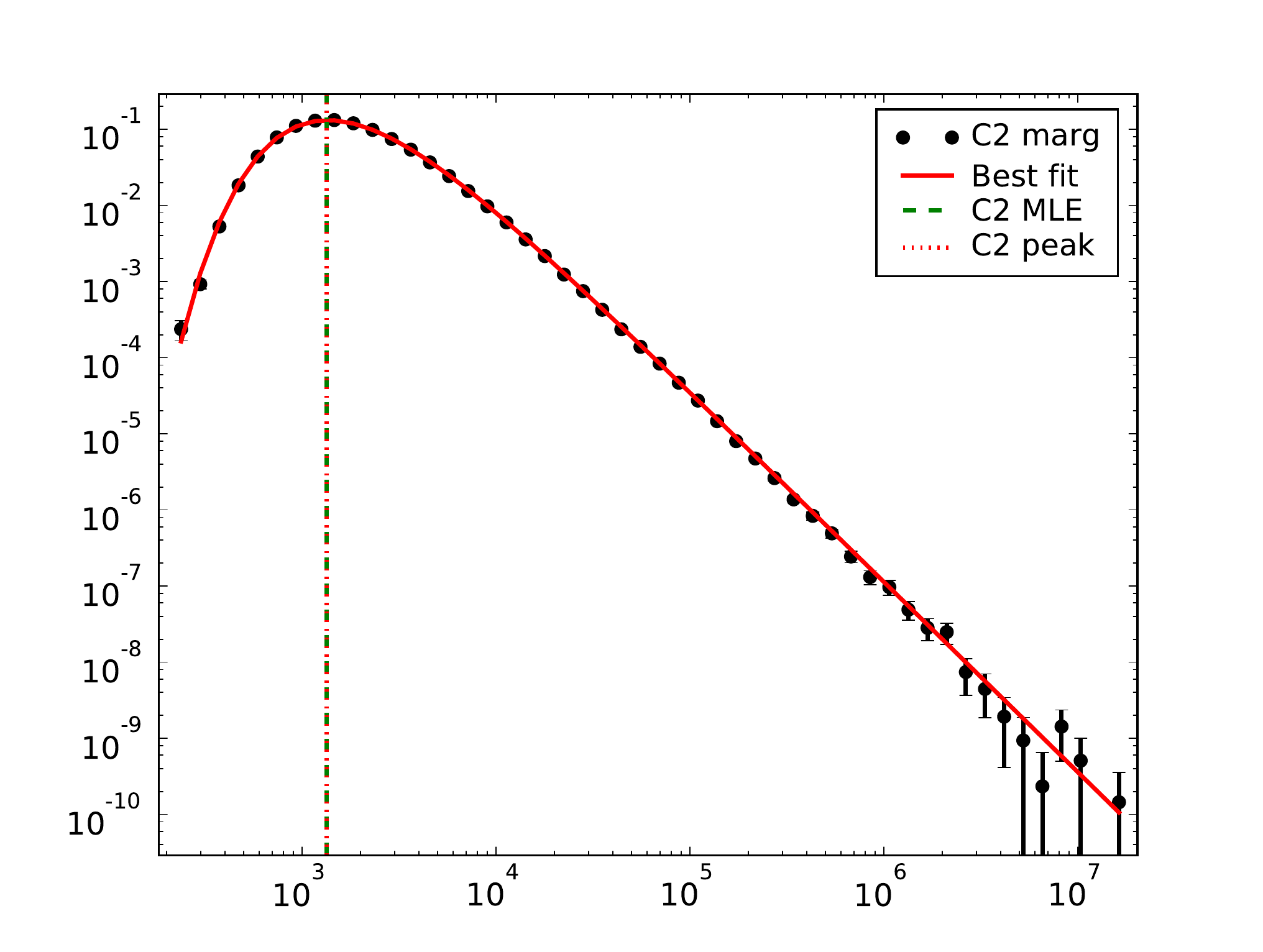}\includegraphics[clip,width=1\columnwidth]{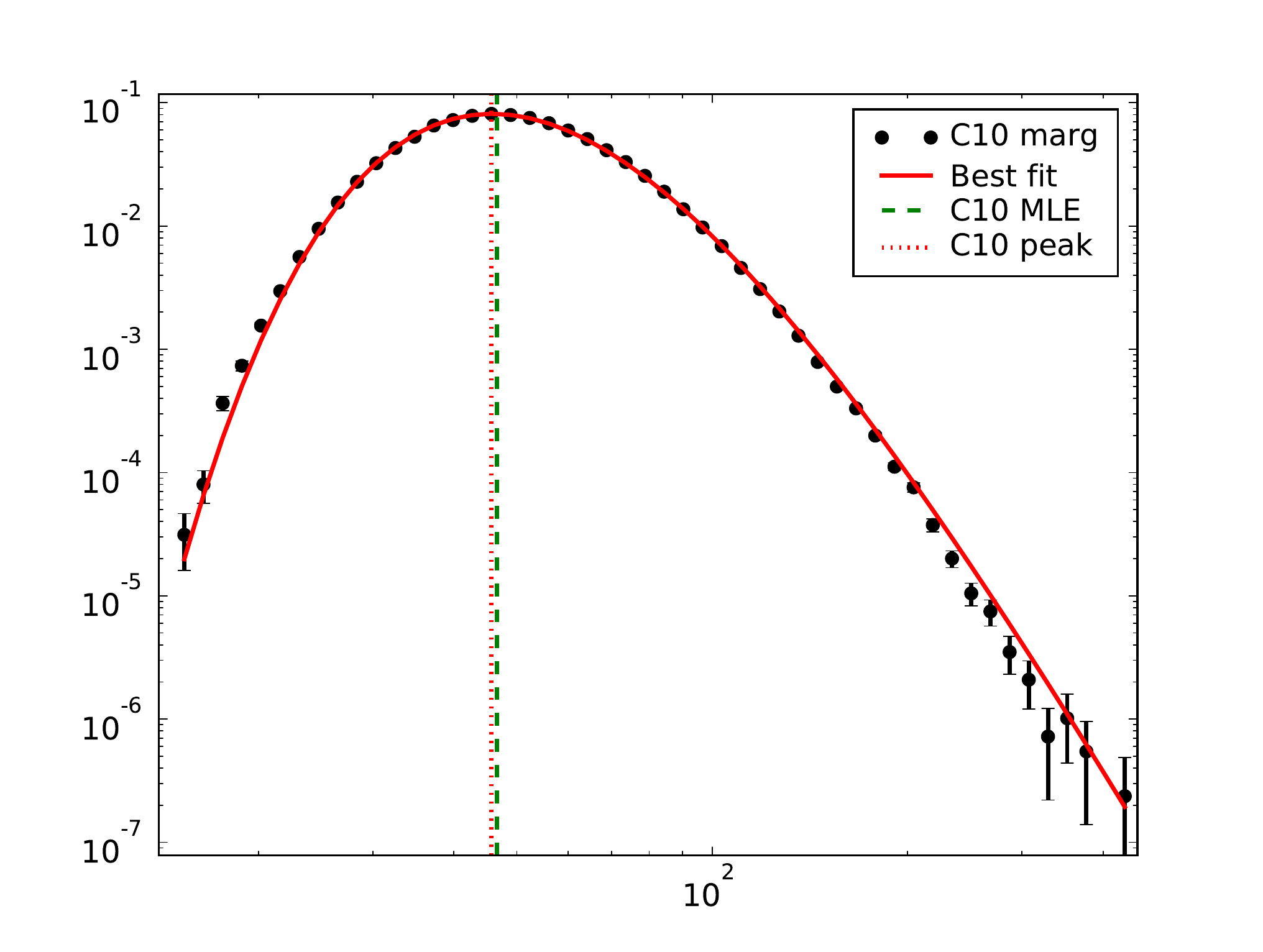}\\
    \includegraphics[clip,width=1\columnwidth]{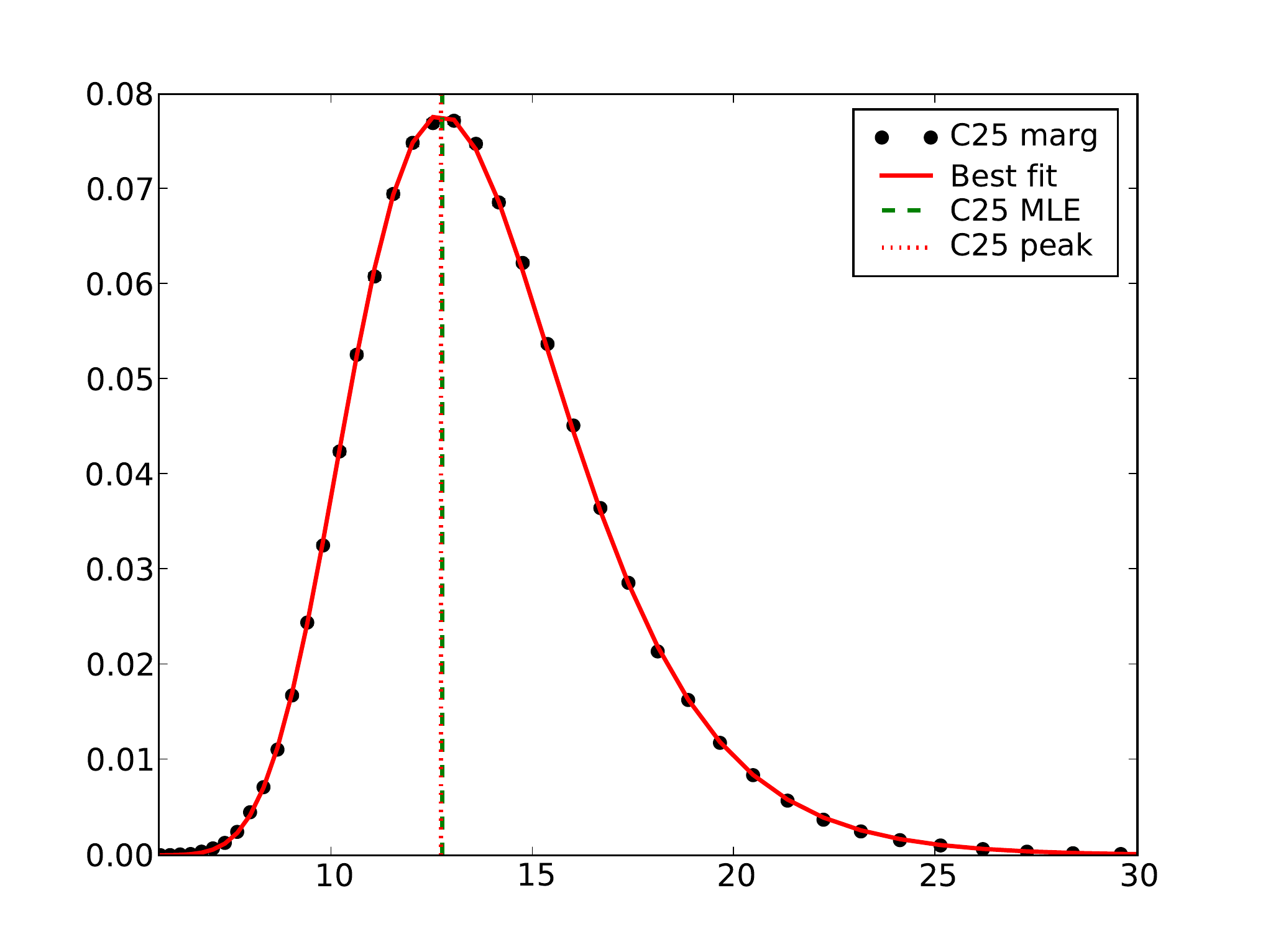}\includegraphics[clip,width=1\columnwidth]{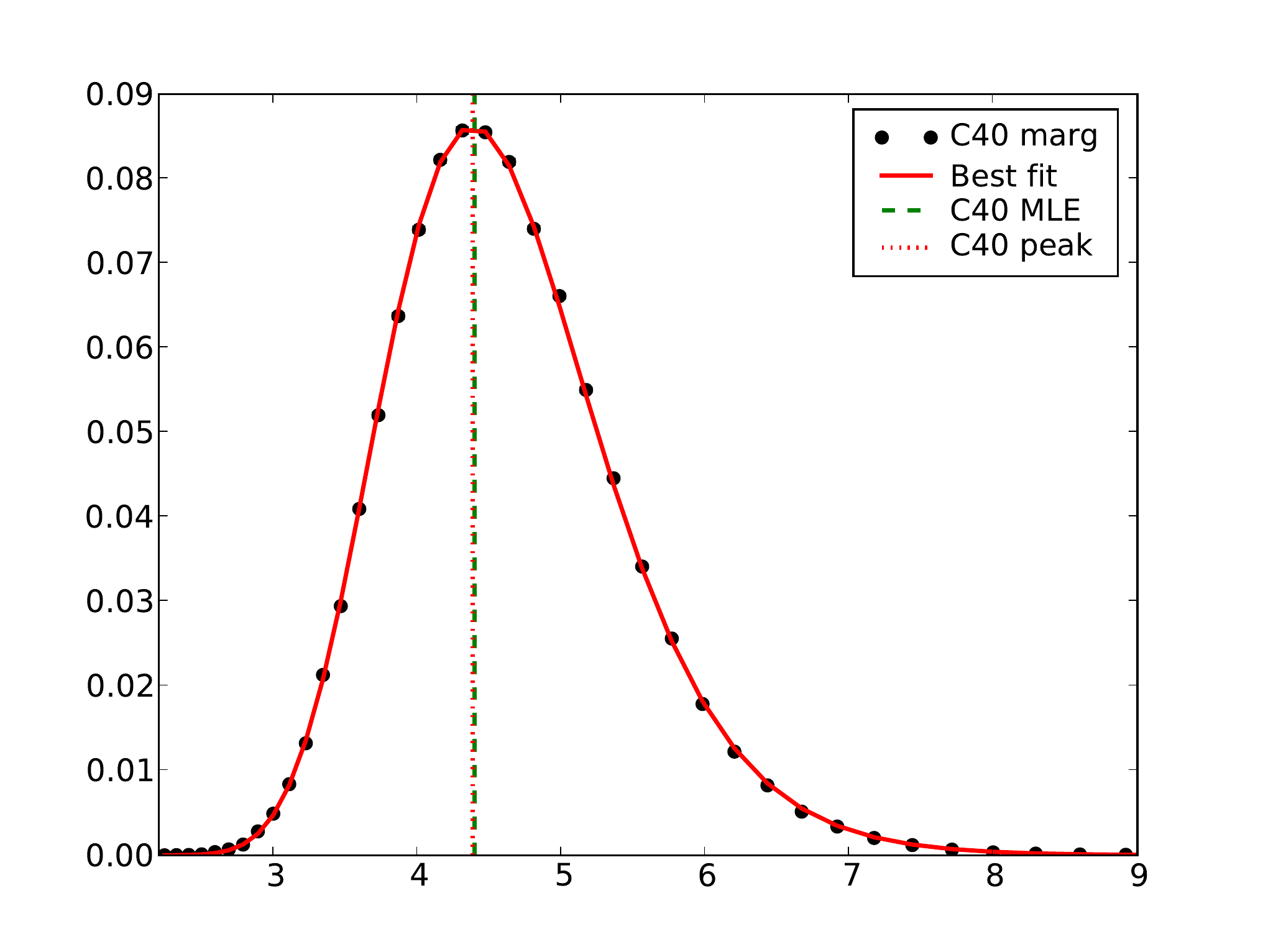} 
    \par
  \end{centering}
  \caption{A few marginalized binned posteriors of the $C_{\ell}$.  
    The red line is the best inverse gamma approximation (including
    binning) obtained using the maximum likelihood estimates, 
    while the black dots are the binned marginal obtained
    on the 500k sample.
    The red short dotted vertical line gives the location of the peak according to
    the approximation, and the green long dotted vertical line shows the
    $C_{\ell}^{ML}$. 
    Top plots are in log-log scale, while bottom plots are in linear
    scale to show the behaviour in the tail and at the peak of the
    marginals. }
  \label{fig:cutsky:marginal}.
\end{figure*}
We build a sample of the posterior of the masked map using the
adaptive importance sampling algorithm described above.  We only
explore $\ell=2$ to $40$, the other modes ($\ell=0,1$ and
$41\leq\ell\leq 48$) being held constant to the ML estimate.

The initial proposal is given by the product of independent inverse
gamma distributions, as described in \ref{sub:adapt:algo}, centered at
$D_{\ell}^{ML}$ with a width given by an effective sky coverage 
equal to $\fsky\times0.98$ to ensure that the initial proposal is 
wide enough.

Only one adaptation step was needed.  It took about $58$min on 80
2~GHz CPUs to produce the first 50k samples (about $6$sec for 
each likelihood evaluation, taking into
account all overheads). The final 500k samples run took $6$~hours 
and $21$ minutes
on 120 2~GHz CPU (about $5.5$sec for each likelihood evaluation,
taking into account all overheads).  The adaptive algorithm behaved
very well: the first step reached $\perplex=0.68$ while the second run
hit $\perplex=0.93$.  This last run had an effective sample size
$ESS=437029$, \textit{i.e.} a ratio $ESS/N=0.874$.

\begin{figure}
  \includegraphics[clip,width=1\columnwidth]{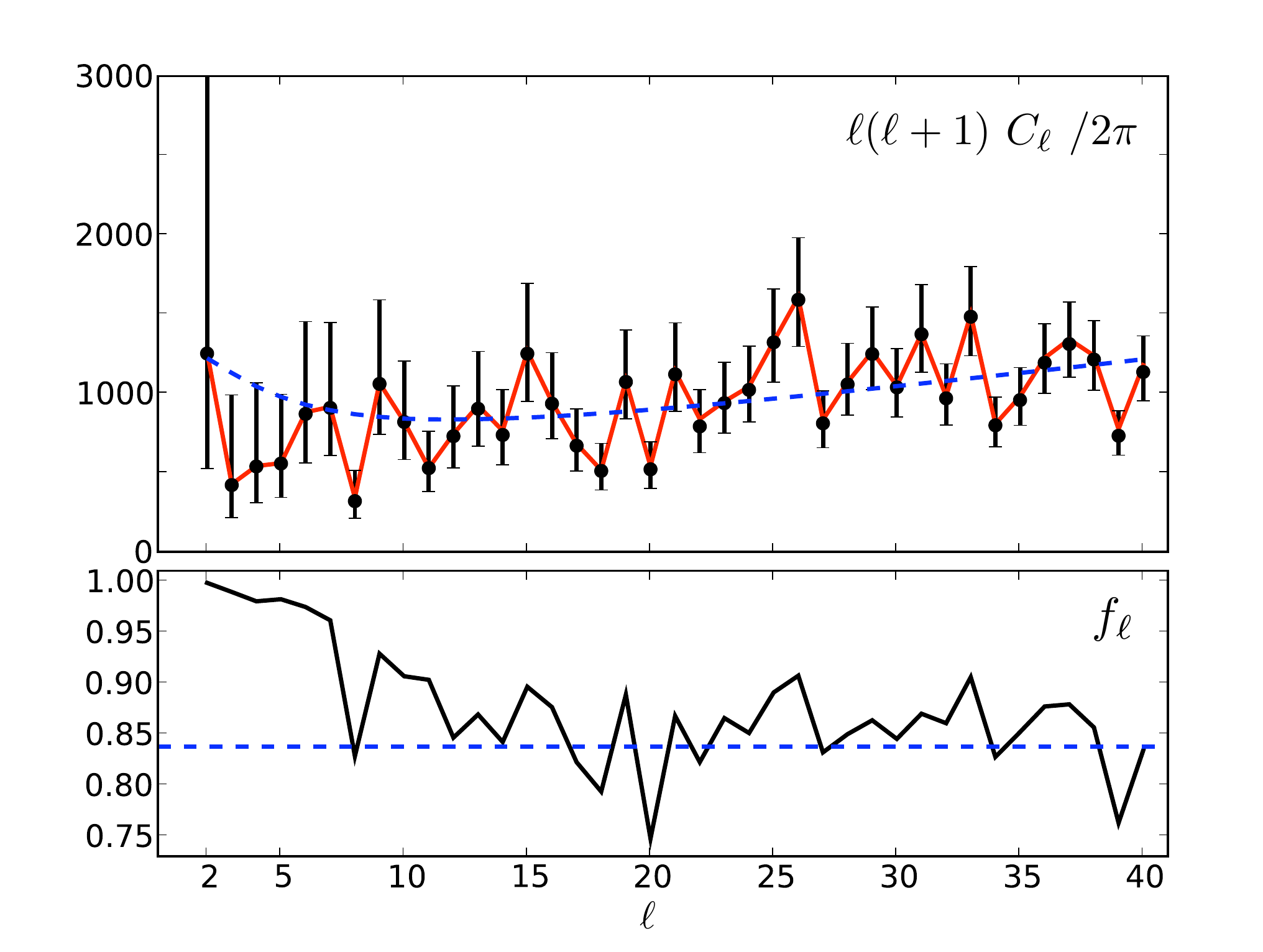}
  \caption{Top panel: angular spectra.  
    Blue dashed line: the power spectrum used to synthesize the map;
    red: the ML estimate $C_{\ell}^{ML}$; black dots: $\peakCl$.  The
    error bars are $68\%$ limits obtained from the inverse gamma fits
    for the marginals.  Bottom panel: Sky coverage $f_{\ell}$.  Black
    line: effective coverage $f_{\ell}$; the blue dashed line shows
    $\fsky=N_{mask} /\Npix$.}
  \label{fig:cutsky:clfl}
\end{figure}

\begin{figure}
  \includegraphics[width=1\columnwidth]{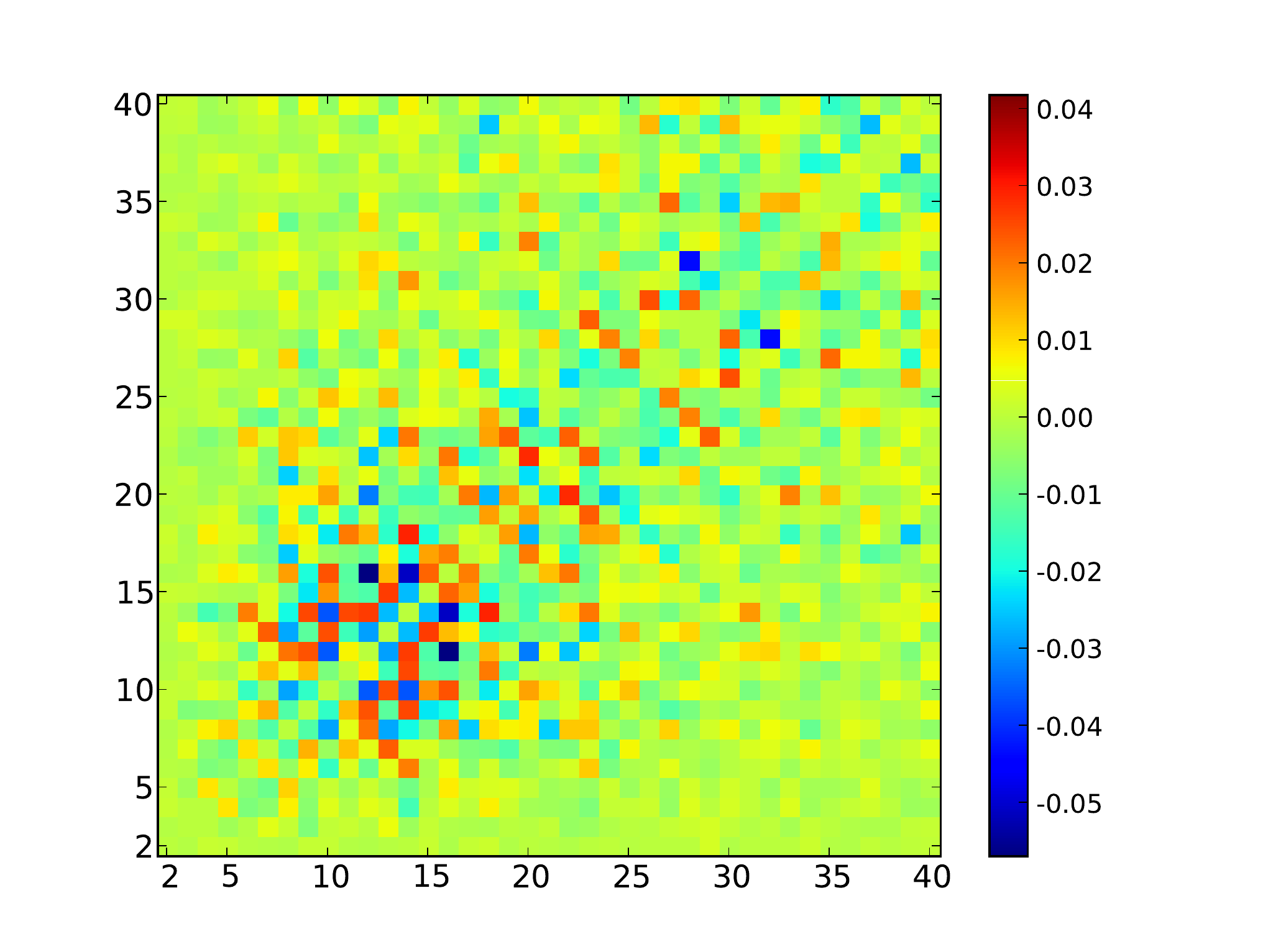}
  \caption{The correlation matrix $V$ for $C_{\ell}$ (see
    Eq.~(\ref{eq:V:def})) with the diagonal removed.  Most of the
    correlation is located around $\ell=12$ and extends only to a few
    neighboring modes.  The correlation is always below the $6\%$
    level.}
  \label{fig:cutsky:cov} 
\end{figure}

Figures.~(\ref{fig:cutsky:marginal})-(\ref{fig:cutsky:cov}) give an
overview of the results.  First, looking at the 1D marginal
distributions, figure~(\ref{fig:cutsky:marginal}) shows a few
marginals ($\pi_\ell$) and their best inverse gamma fits.  The inverse
gamma model is seen to account very well for both the tails and the
mode of the distribution, in line with the high perplexity reached in
the last iteration.  This agreement validates \emph{a posteriori} the
adaptive approach.  On this synthetic map, at least, the marginals
follow closely an inverse gamma distribution.

The peaks of the marginals and an effective sky coverage at multipole
$\ell$, denoted $f_{\ell}$, are obtained by inverting
\begin{eqnarray}
  \alpha_{\ell} & = & \frac{(2\ell+1)}{2}\ f_{\ell}-1\label{eq:alphaL}\\
  \beta_{\ell} & = & \frac{(2\ell+1)}{2}\
  f_{\ell}\left(W_{\ell}\peakCl+N_{\ell}\right),\label{eq:betaL}
\end{eqnarray}
Both quantities are shown in figure (\ref{fig:cutsky:clfl}).  The
$C_{\ell}^{peak}$ and $C_{\ell}^{ML}$ discrepancy is small; it is
below the percent order, albeit with a few modes disagreeing by at
most~$3\%$.  The effective sky coverage, however, is quite different
from $\fsky$.  Its behaviour indicates a transition between scales
that are not affected significantly by the cut, and scales that are
smaller than the cut, so that their deficit of modes is given by
$\fsky$.  Our resolution is probably not good enough to reach this
regime.

One would expect some discrepancy between the $\peakCl$ and the ML
estimate.  Indeed, since the cut induces correlation between scales,
there is no reason for the peak of the posterior to be identical to
the peak of the marginals in each direction.  The small discrepancy
can only be explained by a low level of correlation between the
$C_{\ell}$s, so that the peak of the marginals is close to the joint
peak.
As a first estimate of the correlation, figure.~(\ref{fig:cutsky:cov})
shows the correlation matrix measured on our sample
\begin{equation} 
  \label{eq:V:def}
  [V]_{\ell,\ell'} \equiv \mathrm{Corr} \left( C_{\ell}, C_{\ell'}\right) .
\end{equation}
In this figure, the diagonal of the matrix is removed so as not to
dominate the off diagonal terms.  Those exhibit a pattern below the
$6\%$ level.  Most of the correlation is located around $\ell=12$, and
the correlation seems to extend significantly for about $6$ modes off
the diagonal.

Several checks can be performed to assess the accuracy of this matrix.
First, the effective sample size allows us to estimate the error on
the matrix measurement to be of the order of $0.15\%$, which is well
below the observed correlation pattern.  One can also measure the
correlation matrix on the results of the first iteration of the
adaptive algorithm, which provides us with an independent exploration
of the posterior.  The noise was much higher (with a level, according
to the $ESS$ of this run of about $0.6\%$), but the pattern observed
on figure~(\ref{fig:cutsky:cov}) is easily recovered.  Finally, we checked
on a \emph{full sky} run that no correlation  pattern  is visible.

\subsection{WMAP5 map}\label{sec:samplewmap5}

We perform a similar experiment using the WMAP map distributed along
with the five year WMAP likelihood code found on the Lambda
website~\footnote{\href{http://lambda.gsfc.nasa.gov/}{http://lambda.gsfc.nasa.gov/}}.  The setting is
slightly different, since the window function is a $9.18^{\circ}$
Gaussian beam, cutting much more high frequency power than the window
function~(\ref{eq:archcos}) (see figure~\ref{fig:wlnlcl}). Therefore,
only the range $2\leq \ell\leq 30$ is explored here, with the other
multipole powers held constant at their ML values.
As done in the WMAP likelihood code, a $1\mu$K/pixel noise is added
to the data and to the model.  We take care of adding the specific
noise realization used in the likelihood code.  Indeed, with the beam
used, the signal to noise at $\ell=30$ is only $\sim14$ and our tests
have shown a small dependency of the value of the higher $C_\ell$s on
the noise realization.

As in the previous run, only one adaptation step turns out to be
needed.  It took $32$ minutes on 120~CPUS for 50k samples, 
while the second and final
run produced 500k samples in $5$ hours and $19$ minutes.  
The first iteration reached
$\perplex=0.48$, the second one $\perplex=0.96$ and an effective
sample size $ESS=457600$ ($ESS/N=0.92$).

\begin{figure}
  \includegraphics[width=1\columnwidth]{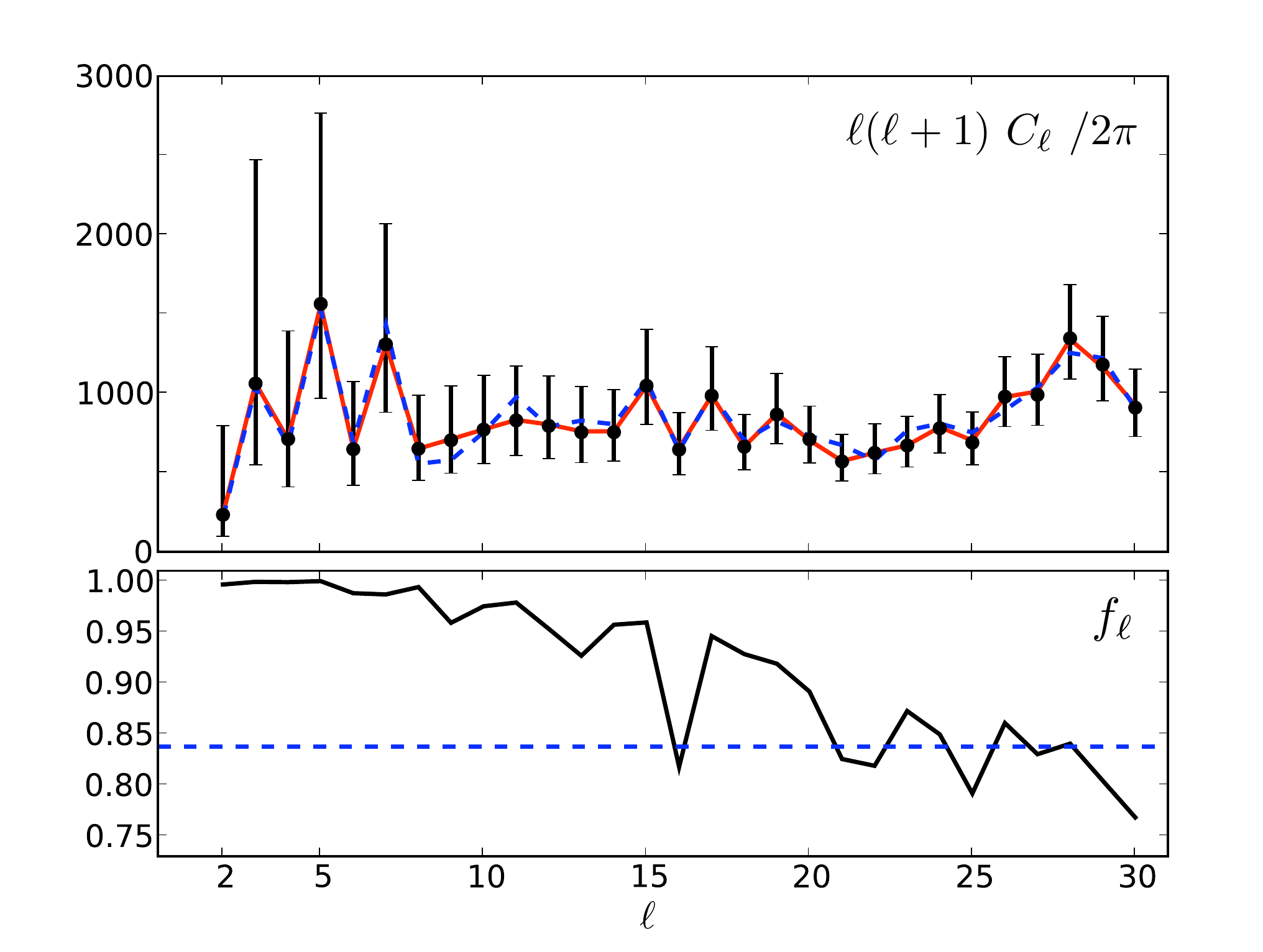}
  \caption{Same as figure~\ref{fig:cutsky:clfl} for the WMAP5 data set.
    The dashed blue on the top panel of the top panel line now is the
    WMAP empirical spectrum.}
  \label{fig:WMAP5:clfl}
\end{figure}

\begin{figure}
  \includegraphics[width=1\columnwidth]{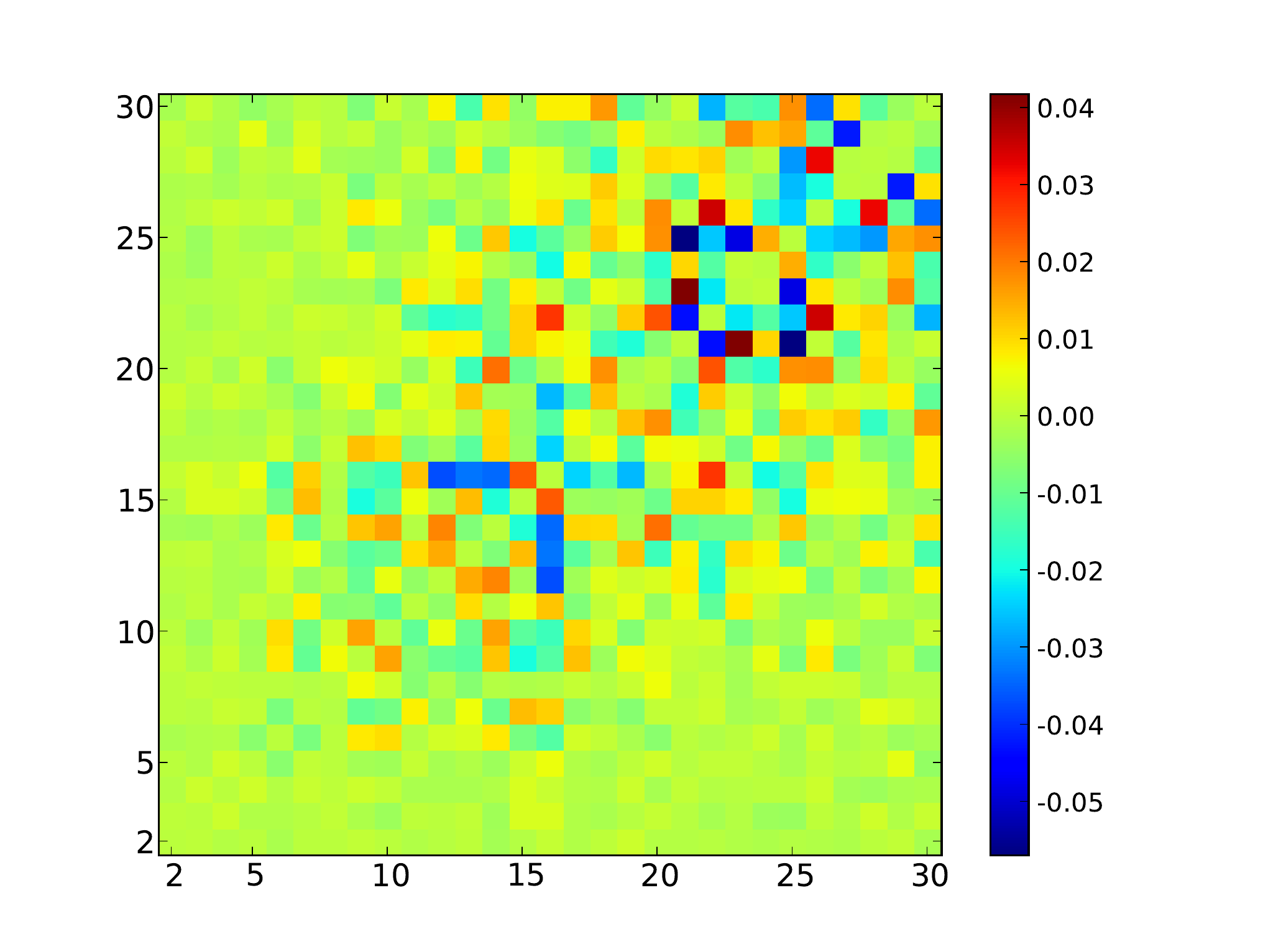}
  \caption{Same as figure~\ref{fig:cutsky:cov} for the WMAP5 data set.}
  \label{fig:WMAP5:cov}
\end{figure}

The results are generally similar to those reported in
section~\ref{sec:synthmap}.  We do not show more 1D marginal plots,
but present the recovered $\peakCl$ and $f_{\ell}$
(figure~\ref{fig:WMAP5:clfl}), as well as the correlation matrix
(figure~\ref{fig:WMAP5:cov}).
The $\peakCl$ and the ML estimates are somewhat similar to the WMAP5
power spectrum, with a small discrepancy also observed by \cite{Eriksen:2006p3364}
using Gibbs sampling  and in \cite{Rudjord:2008p4864} (zooming on their
figure 5).  At any rate, the discrepancy is always within the $C_{\ell}$
error bars.

The effective coverage $f_{\ell}$ is similar to the one reported in
section~\ref{sec:synthmap}, with a transition from $1$ to
$\fsky$ but differs in some details, indicating that it is not only a
function of the mask, but also of the actual data set.

Finally, figure~(\ref{fig:WMAP5:cov}) shows the correlation matrix.  It
exhibits structures similar to those in figure~(\ref{fig:cutsky:cov}).
As for the $f_{\ell}$, the differences between
figures~\ref{fig:WMAP5:cov} and \ref{fig:cutsky:cov} indicate that the
correlation matrix does not depend only on the mask.

\section{Approximating the low-$\ell$ likelihood}\label{sec:lapprox}

For both data sets considered in previous section, the posterior
distribution of the total angular spectrum $D_\ell$ revealed similar
and striking features: the marginals are very well approximated by
inverse gamma distributions and there is a weak correlation between multipoles
(below the $10\%$ level).
Since we used a flat prior, these findings suggest that a copula
approximation to the likelihood should be quite accurate (in addition
to being fast, by design).
This approach is somewhat similar to what has been proposed by
\cite{BJK:2000ApJ...533...19B} and implemented at low-$\ell$ in
\cite{Rudjord:2008p4864} and at high-$\ell$ in
\cite{Hamimeche:2008p3608}.  It differs in that, instead of offset log
normal (as in \cite{BJK:2000ApJ...533...19B}), spline approximation
\citep{Rudjord:2008p4864} or Taylor expansion inspired approximation
\citep{Hamimeche:2008p3608}, we use inverse gamma cumulative functions
for Gaussianization.

\subsection{Copula approximation}\label{sec:copula-approximation}

A good approximation formula must at least reproduce the inverse gamma
marginals, and the observed level of correlation.  A generic way of
building multivariate distributions with specified marginals and some
correlation is provided by \emph{copula models}~\citep{sklar59}.

\medskip\noindent\textbf{The copula model.}
Denote $\Normal^{(d)}(\cdot; \mu, M)$ the $d$-variate Gaussian density
with mean $\mu$ and covariance matrix $M$.
Consider a set of zero-mean unit-variance Gaussian variables
$G_{\ell}$ with density $\Normal^{(d)}(G_{\ell};0, M_{G})$ where $M_G$
has only $1$'s on the diagonal and possibly non-zero off diagonal terms.
Consider those transformed variables $D_\ell=D_\ell(G_\ell)$ which
have an inverse gamma distribution with parameters $\alpha_\ell$ and
$\beta_\ell$, that is, $G_\ell$ and $D_\ell$ are related by
\begin{equation} 
  \label{eq:repar:gauss:integ}
  \Normal(G_{\ell};0,1)\
  dG_{\ell}=i\Gamma(D_{\ell};\alpha_{\ell},\beta_{\ell})\  dD_{\ell} .
\end{equation}
The distribution of $D_\ell$ is then easily seen to be
\begin{equation}
  \label{eq:approx:def}
  \tilde{\pi}(D_{\ell})
  \equiv
  \prod_{k}i\Gamma(D_{k};\alpha_{k},\beta_{k}).
  \,
  \frac
  {         \Normal^{(d)} (G_{\ell};0, M_{G})}
  {\prod_{k}\Normal^{(1)} (G_{k}   ;0, 1)}
  .
\end{equation}
Distribution~(\ref{eq:approx:def}) is called the \emph{copula
  approximation}.
It belongs to a parametric model with $2d+d(d-1)/2$ parameters: each
of the $d$ multipoles requires a pair $(\alpha_\ell,\beta_\ell)$ for
the marginal distribution and the correlation matrix $M_G$ depends on
$d(d-1)/2$ free parameters.

\medskip\noindent\textbf{Two properties.}
Probability distributions of the form~(\ref{eq:approx:def}) enjoy two
nice properties which readily follow from their construction.
First, the marginal distribution of each $D_\ell$ remains an inverse
Gamma regardless of the correlation level (which is independently
controlled by the matrix $M_G$).
Second, marginalization over any subset of $D_\ell$ is readily
achieved by removing the corresponding rows and columns of matrix
$M_G$.

\medskip\noindent\textbf{Gaussianization.}
Evaluating the copula density~(\ref{eq:approx:def}) requires explicit
Gaussianization, that is mapping $D_\ell$ to $G_\ell$.  This is easy
since relation~(\ref{eq:repar:gauss:integ}) implies that
\begin{equation}
  G_{\ell}
  \equiv 
  \mbox{c}N^{-1}( \mbox{c}i\Gamma(D_{\ell};\alpha_{\ell},\beta_{\ell})),
  \label{eq:repar:gauss:def}
\end{equation}
where $\mbox{c}i\Gamma(\cdot; \alpha, \beta)$ denotes the cumulative
distribution function (CDF) of the inverse gamma distribution and
$\mbox{c}N^{-1}$ is the inverse CDF (or quantile function) of the
standard normal distribution, sometimes called the \textit{probit}
function.
The former is 
\begin{displaymath}
  \mbox{c}i\Gamma(x;\alpha,\beta)\equiv\int_{0}^{x}i\Gamma(t;\alpha,\beta)\
  \mbox{d}t=\Gamma\left(\alpha,\beta/x\right)/\Gamma(\alpha).
\end{displaymath}
while the latter, if missing from a statistical library, can be
computed as $\mbox{c}N(x)^{-1} = \sqrt{2} \mbox{erf}^{-1}(2x-1)$ with
$\mbox{erf}(y) = \frac{2}{\sqrt{\pi}}\int_{0}^{y}\exp(-t^{2})\
\mbox{d}t$.

\medskip\noindent\textbf{Speed.}  
Copula evaluation is very fast.
Using a custom code to compute the inverse error function, and the
free GSL library%
\footnote{\href{http://www.gnu.org/software/gsl/}{http://www.gnu.org/software/gsl/}%
} for the gamma and cumulative gamma distribution, we can compute
about $18000$ samples per second while the pixel-based likelihood
needs about $5.5$ seconds per sample on the same computer within the
same setting (\textit{i.e.} same overheads).  
Moreover, one can also sample directly from the copula by first
drawing $G_{\ell}$ under to their multivariate Gaussian distribution
and then invert Eq.~(\ref{eq:repar:gauss:def}) to get the $D_{\ell}$
values.

\medskip\noindent\textbf{Learning the copula model.}
Learning the $2d+d(d-1)/2$ parameters of a copula models from
(importance) samples of $D_\ell$ is straightforward.  In a first step,
one estimates for each $\ell$, the inverse gamma parameters
$(\alpha_\ell, \beta_\ell)$ by maximum likelihood (see
appendix~\ref{sec:mlegamma}).
In a second step, the samples are Gaussianized via
eq.~(\ref{eq:repar:gauss:def}) using the estimated values of
$(\alpha_\ell,\beta_\ell)$.  Finally, matrix $M_G$ is plainly
estimated as the sample correlation matrix of the Gaussianized
samples.

\medskip\noindent\textbf{Significance of correlation.}
Given a copula model $\tilde\pi$ with correlation matrix $M_G$, there
is a simpler copula model with the same marginals but without
correlation, that is, with $M_G=I$.  This model is denoted
$\tilde\pi_0$ and called the \emph{uncorrelated model} which, of course,
is not as accurate as $\tilde\pi$.  
Since $\tilde \pi_0$ and $\tilde \pi$ are Gaussian distributions, the loss can be quantified exactly thanks to a Pythagorean property of
the KLD which yields
\begin{equation}\label{eq:pythasym}
  K(\pi | \tilde \pi_0 )  = K(\pi | \tilde{\pi})  + K(\tilde{\pi} | \tilde\pi_0 ) .
\end{equation}
It shows that the mismatch $K(\pi | \tilde \pi_0)$ of the uncorrelated
approximation to the posterior is larger than the mismatch
$K(\pi|\tilde \pi)$ of the regular copula by a positive term
$K(\tilde{\pi}|\tilde\pi_0 )$.  This term can
be computed in closed form:
\begin{equation}
  \label{eq:kullcorr}
  K(\tilde{\pi}|\tilde \pi_0)=-\frac12 \log\det M_G
\end{equation}
which is positive unless $M_G=I$ and readily gives a measure of the
price to pay for ignoring correlation.

\subsection{Validation : first results  }

We first look at some self-consistency results when learning a copula
model from the importance samples obtained from the WMAP data set
discussed in section~\ref{sec:samplewmap5}.

\medskip\noindent\textbf{High perplexity.}  The first important thing
to report is that, on the perplexity scale, the copula approximation
is remarkably good: we reach $\perplex(\pi|\tilde{\pi})=0.99$ on a
500k simulation sample using estimates of the $\peakCl$, $f_{\ell}$
and $M_G$ obtained on the same sample.
As a simple cross-validation test, we split the 500k sample into two
subsets of equal size, re-estimate the copula parameters on the first
subset and compute the perplexity using the second subset. We find a
negligible decrease in perplexity of about $5\times10^{-4}$.

Thus, the copula approximation appears to work extremely well on this
data set.  Still, one should look further than a single number.  This
section looks into more details of the approximation.

\begin{figure}
  \includegraphics[width=1\columnwidth]{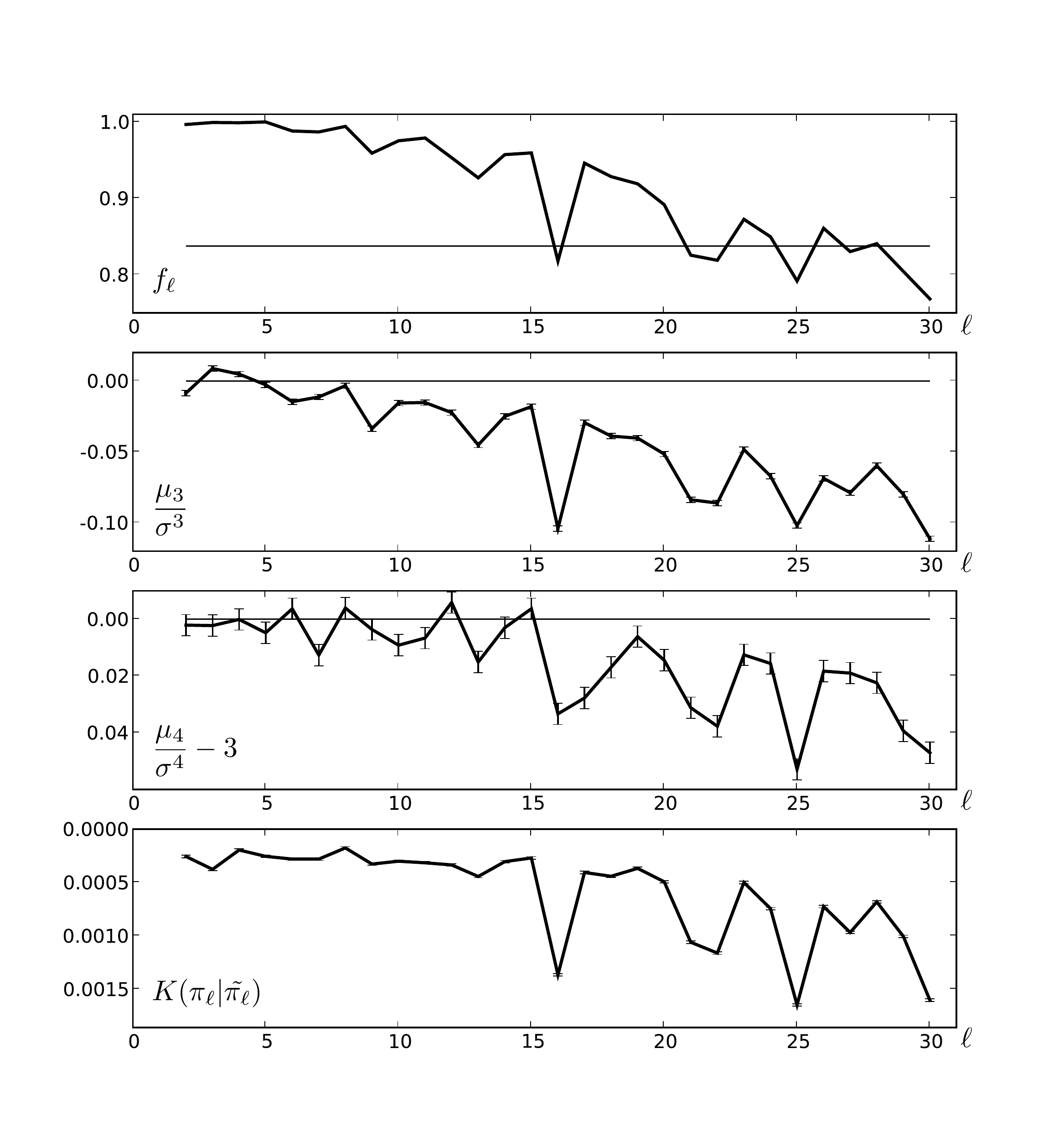}
  \caption{$f_{\ell}$, cumulants and Kullback divergence of 
    $G_{\ell}$ exhibit some correlation.  From top to bottom, $f_\ell$
    (and $\fsky$), skewness, kurtosis and Kullback divergence between
    the marginals and standard normal.
    Note that the Kullback divergence is estimated from an histogram.
    The last two panels have their ordinates downwards to better
    show the correlation.  Error bars are measured on 500 Gaussian
    simulations of size $ESS$ $(=457600)$}
  \label{fig:flcor}
\end{figure}

\medskip\noindent\textbf{Gaussianization.}
Even though the marginals were found to be well approximated by
inverse gamma distributions, the Gaussianized importance samples may
show some small hints of\ldots non Gaussianity.
Indeed, for each $G_\ell$, we computed the skewness, the kurtosis and
the Kullback divergence to a standard Gaussian.  See
figure~\ref{fig:flcor} for the WMAP data set (a similar plot can be
obtained on the other data set).
The plot shows a small deviation from Gaussianity showing that the
target densities are not \emph{exactly} inverse gamma distributed.
In addition, those non Gaussian indicators degrade with $\ell$ and are
correlated with $f_\ell$.  Since the latter measures deviation from
the full-sky case, this is not unexpected.

\begin{figure*}
  \begin{tabular}{ccc}
    \includegraphics[width=0.63\columnwidth]{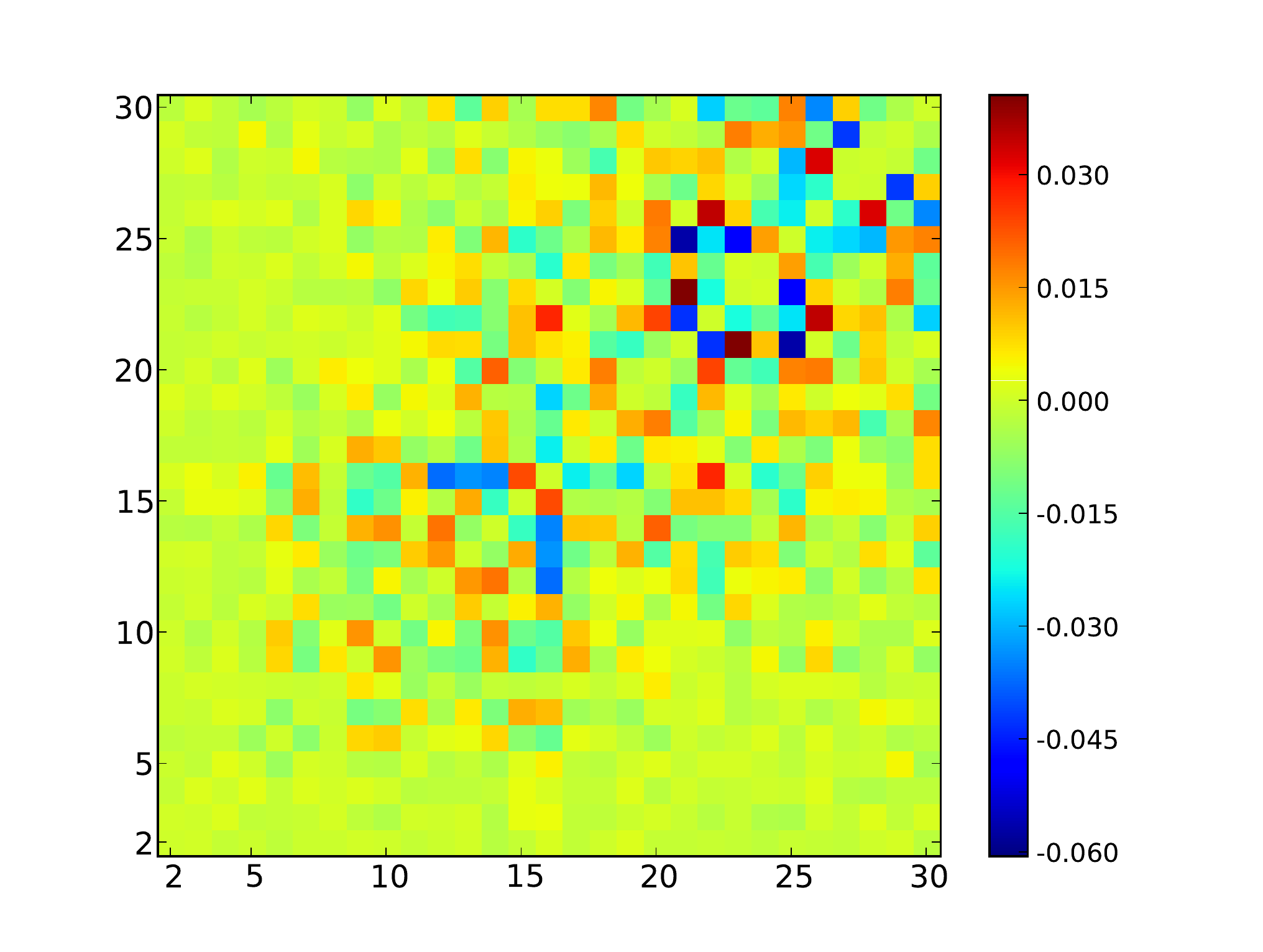} 
    & \includegraphics[width=0.63\columnwidth]{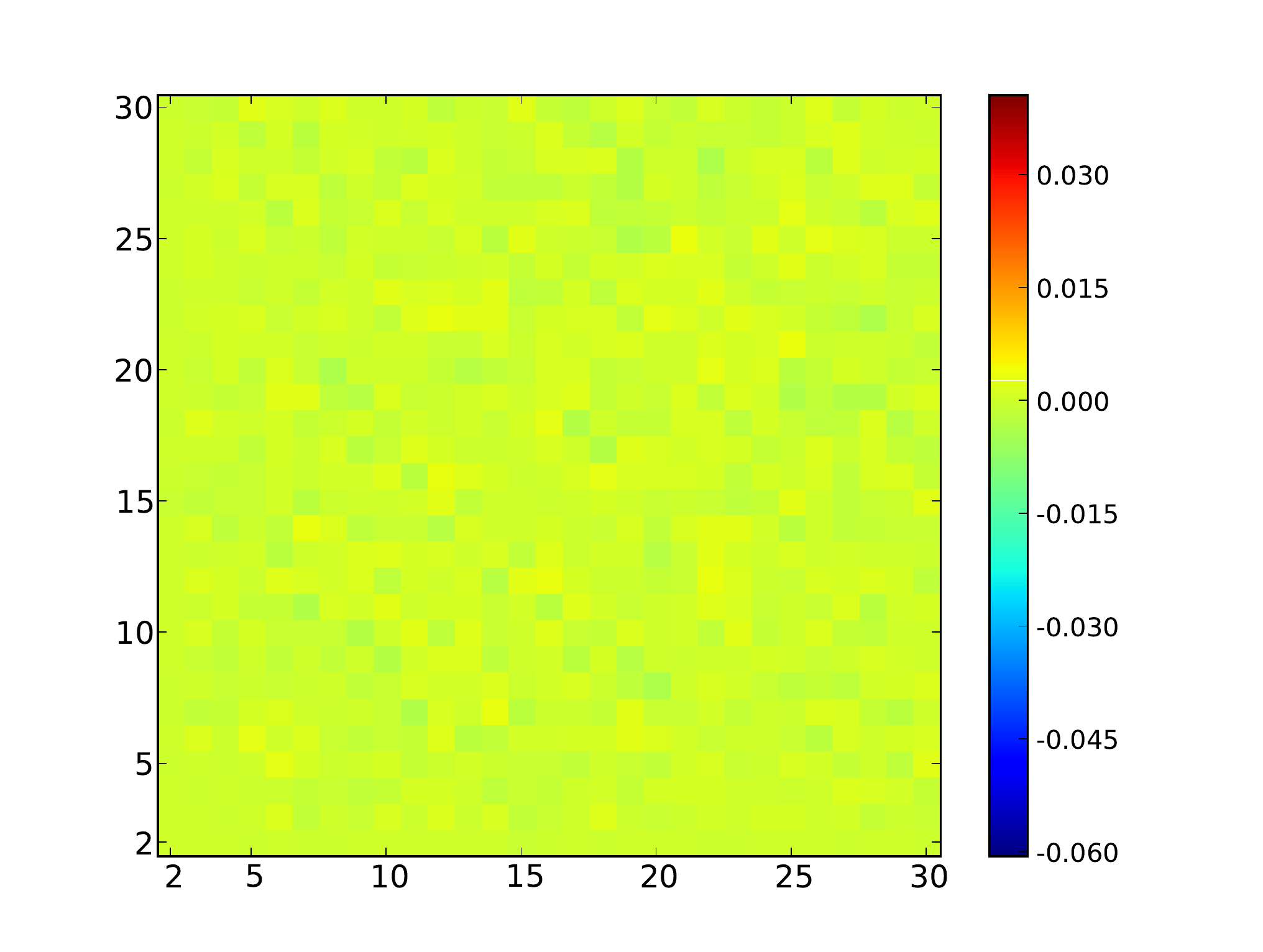} 
    & \includegraphics[width=0.63\columnwidth]{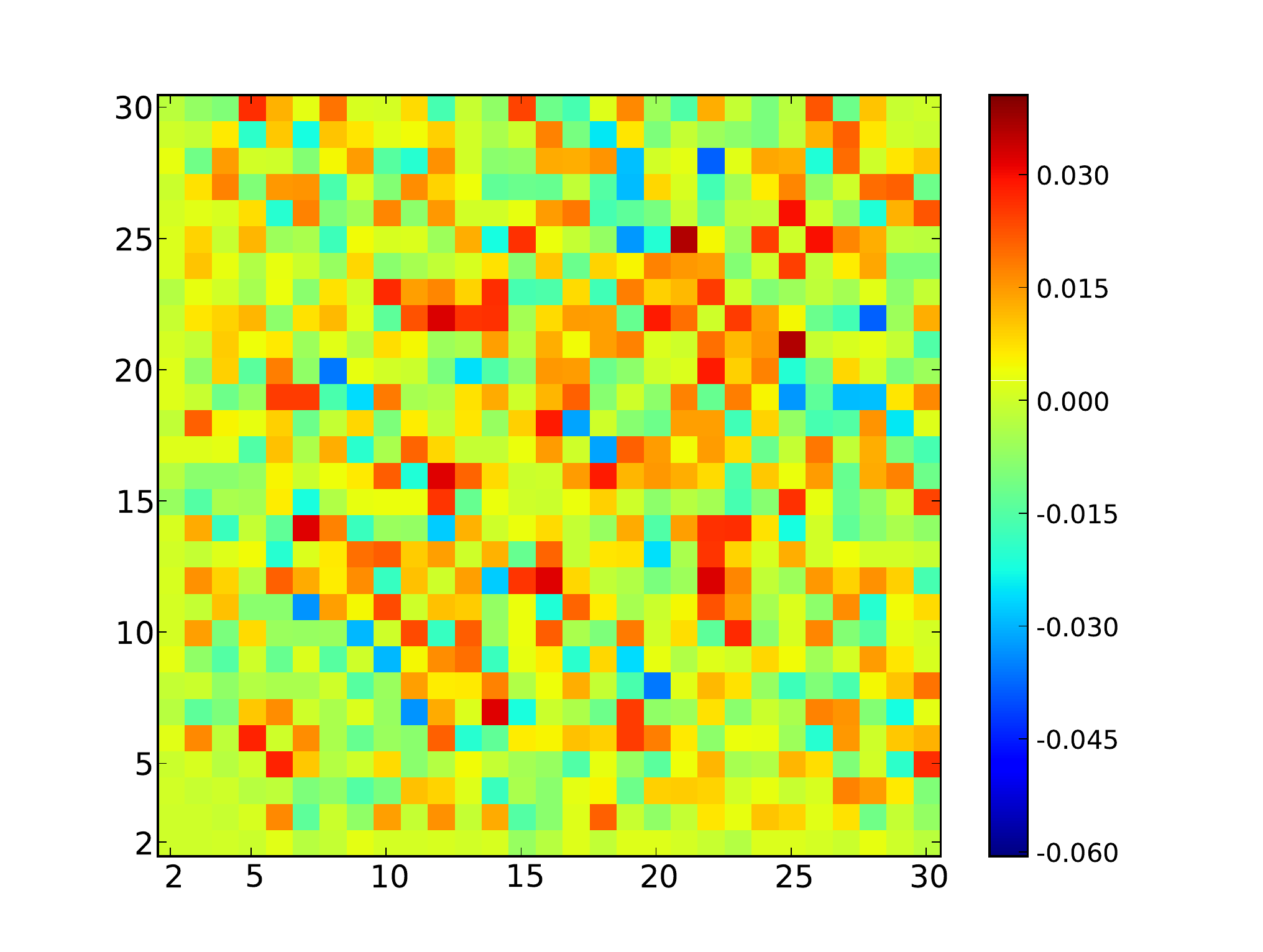} 
  \end{tabular}
  \caption{
    Left panel: $V$; center panel: $V-\tilde{V}$; right panel:
    $10\times(V-\tilde{V})$.  All panels share the same color scale.
    Matrices $V$ and $\tilde V$ have been obtained on the same
    importance sample with appropriate weights.}
  \label{fig:diff:corrBis} 
\end{figure*}

\medskip\noindent\textbf{Correlation matrices.}
By design, the copula correctly predicts the correlation matrix of the \emph{Gaussianized} variables
but it is not necessarily accurate as a predictor of the correlation
matrix $V$ of $C_\ell$.
Here, we check that $V$ is well predicted by the covariance matrix of
the copula model, denoted $\tilde V$.
Matrix $V$ is estimated as described before (based on an importance
sample); matrix $\tilde V$ is obtained from the same importance samples,
re-weighed by $\tilde \pi/\pi$.
The results are displayed on figure~\ref{fig:diff:corrBis} and show an
excellent agreement, with small and evenly distributed errors.

\subsection{Perplexities.}

We briefly report on the relative perplexity and Kullback divergence
between the posterior and its approximations on the WMAP5 data set.
Some results are reported in table~\ref{tab:perp}.
\begin{table}
  \centering
  \begin{tabular}{llr}
    Approximation                     & Perplexity  & Kullback ($\times 10^{-3}$)\\     \hline
    Copula $\tilde\pi$                & 0.991 & $8.6$ \\
    Uncorrelated copula $\tilde\pi_0$ & 0.965 &  $35.2$ \\
    Uncorrelated last run             & 0.956 & $45.0$\\
    Naive $\tilde\pi_\mathrm{naive}$   & 0.779 & $249.6$  \\
    LogNormal                         & 0.191 & $1655.3$  \\ \hline
  \end{tabular}
  \caption{Perplexities.  See text.}
  \label{tab:perp}
\end{table}
Since the Gaussianized variables were found to be weakly correlated,
it may be tempting simply to ignore this correlation and to resort to
the \emph{uncorrelated} approximation $\tilde{\pi}_{0}$ defined at
sec.~\ref{sec:copula-approximation}.
In this case, the fit is slightly degraded: we measure
$\perplex(\pi|\tilde{\pi}_{0}) = 0.97$, in line with the perplexity
obtained after the last step of the adaptive importance run
($\perplex=0.96$, sec.~\ref{sec:samplewmap5}) showing that the
determination of $\peakCl$ and $f_{\ell}$ is only marginally improved
by the 500k simulation.
The contribution of correlation to the quality of the fit is given on
the Kullback scale by the Pythagorean
decomposition~(\ref{eq:pythasym}).  Numerical evaluation by Monte
Carlo integration gives, term-to-term:
\begin{equation}\label{eq:pythnum}
  35.18 \ 10^{-3} \approx 8.61 \ 10^{-3} + 27.3 \ 10^{-3}.
\end{equation}
This is only an approximate equality because of MC errors.  The last
term was also evaluated using eq.~(\ref{eq:kullcorr}), yielding $27.5\
10^{-3}$.
These values show that correlation accounts for most part of the
mismatch in the sense that $ K(\pi | \tilde{\pi}) \approx \frac13
K(\tilde{\pi} | \tilde\pi_0 )$.

Those results can be compared to the naive approximation used as the
initial proposal in our adaptive importance sampling runs, that is,
the copula approximation $\pi_\mathrm{naive}$ with $C_{\ell}^{ML}$,
$\fsky$ and ignoring the correlation.
It gives a perplexity of $\perplex(\pi|\tilde{\pi}_{\rm naive})=0.76$
corresponding to a huge increase in Kullback divergence.

Finally, we compute, as a comparison baseline, the perplexity of the
classical offset log-normal
approximation~\citep{BJK:2000ApJ...533...19B}.  The estimation of the
curvature at the peak is easily derived from $f_{\ell}$. The
perplexity goes down to $\perplex=0.2$ for that approximation.

\subsection{Validation : pseudo-cosmological parameters}

We now compare several likelihood functions via their impact on
estimation of (pseudo) cosmological parameters from WMAP data.
Since only the low $\ell$ part of the spectrum is considered, only a
few cosmological parameters can be fitted.  We choose to perform our
comparisons using a simple model with only two parameters, amplitude
and spectral index, that is, we consider
\begin{equation}
  \tilde{C}_{\ell}
  \equiv 
  C_{\ell}^{\rm ref}\times  A\left(\frac{\ell}{\ell_{0}}\right)^{n},
\end{equation} 
where $C_{\ell}^{\rm ref}$ is a reference angular spectrum (here the
WMAP1 best fit spectrum) and where the relative amplitude $A$ and the
relative spectral index $n$ are our pseudo-cosmological parameters.
The reference power spectrum being a fit on a broader range of
  multipoles, the posterior of $(A,n)$ is not centered at $(1,0)$.

\begin{figure}
  \includegraphics[width=1\columnwidth]{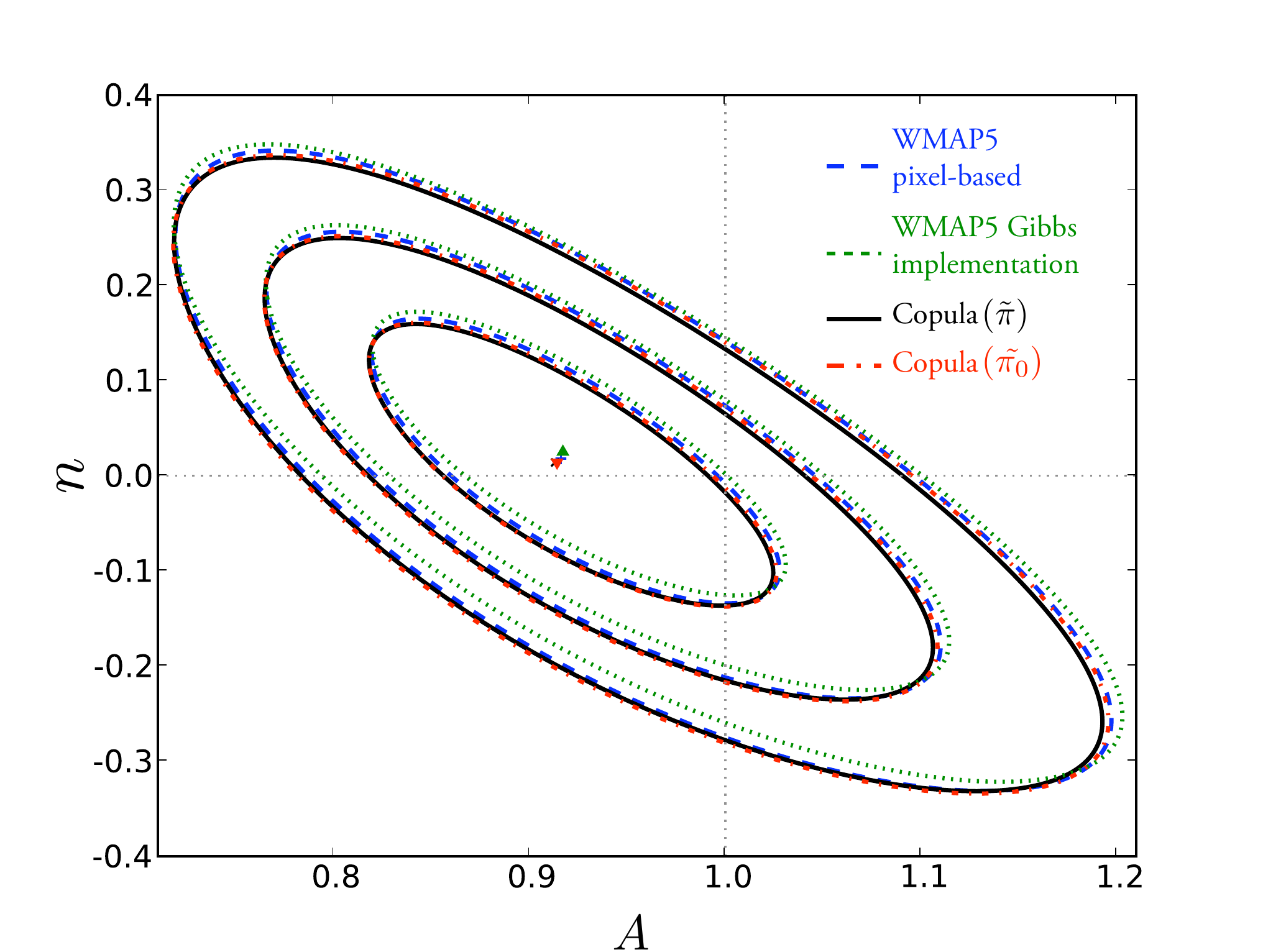}\\
  \includegraphics[width=1\columnwidth]{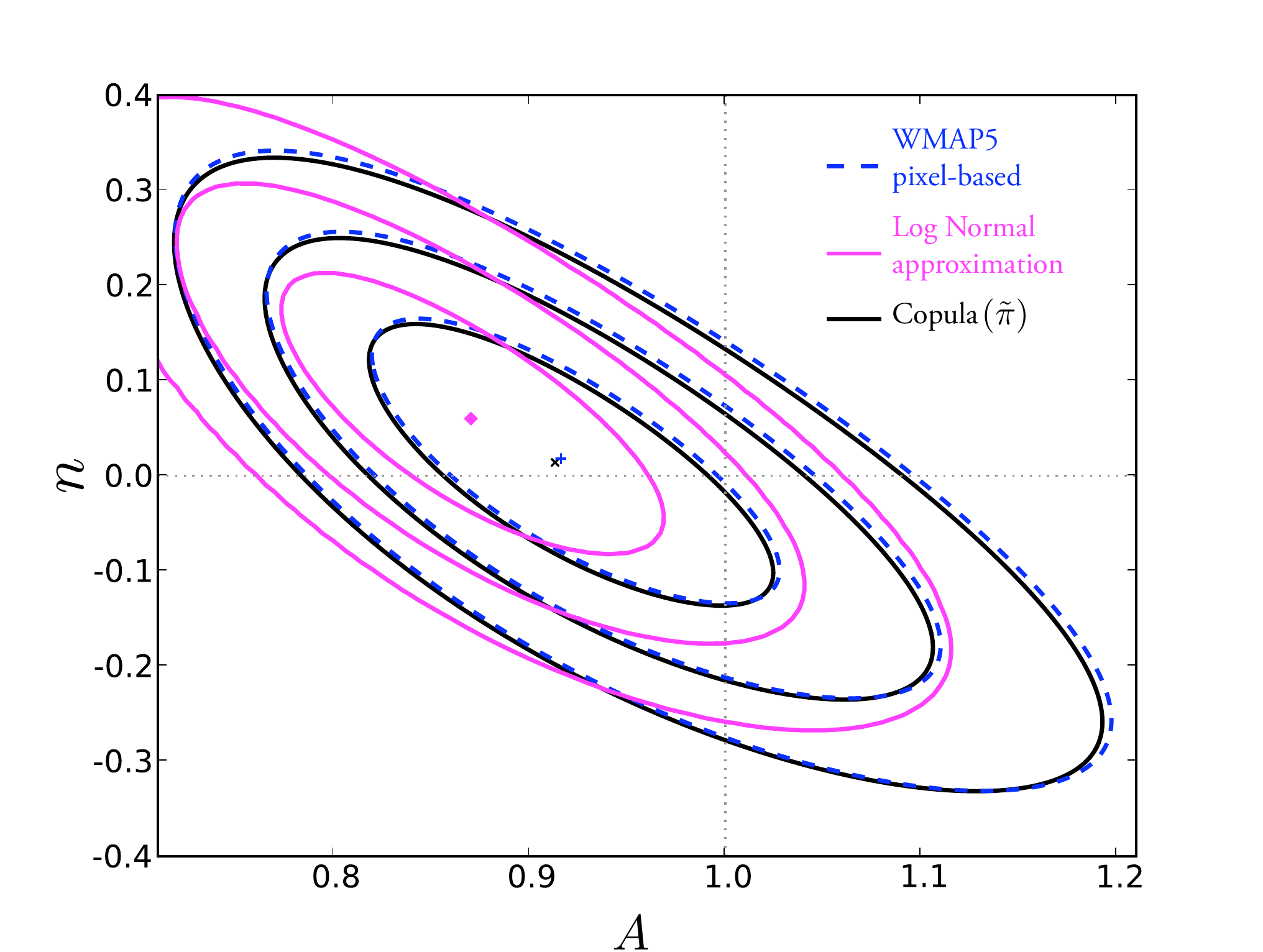}
  \caption{Posterior distribution for $(A,n)$ using different
    likelihood approximations.  
    Both panels:  the dashed blue line shows official WMAP5 likelihood
    code and the black solid line shows the copula approximation $\tilde \pi$. 
    Top panel: green dotted line is the Gibbs implementation included
    in the official code, the red dash-dotted line is the copula
    approximation ignoring correlations, $\tilde \pi_0$.  
    Bottom panel: solid magenta line is the log-normal
    approximation. 
    The colored symbols mark  the peak of  each posterior.}
  \label{fig:anplots}
\end{figure}

Figure~\ref{fig:anplots} shows the $1,2$ and $3\,\sigma$ contours and
the peak position for different likelihood approximations.
The top panel presents a comparison between the WMAP5 likelihood code,
used both in pixel based and Gibbs mode \citep{Dunkley:2008p3305}, and
copula approximations with or without correlations
(i.e. $\tilde \pi$ and $\tilde \pi_0$).  They all appear
to be in remarkably good agreement.  The small
discrepancies in the contour curves (which are smaller than
the grid step size) are much smaller than the width of the mode.  The
peaks of the copula approximations and of the Gibbs approximation are
very slightly displaced compared to the official WMAP5 results, at a
distance of the order of the step size of the grid on which
likelihoods are evaluated.
The bottom panel presents a comparison with the log-normal
approximation described in previous chapter.  As expected, the quality
of that last approximation is poor, with a deviation of the best fit
$(A,n)$ of the order of $\sigma/4$.  Nonetheless, the areas of the
$1,2$ and $3\sigma$ regions are similar, probably because these areas
are mostly controlled by the values of $f_{\ell}$.

\section{Conclusion}

Using an adaptive importance sampling algorithm, we explored the
low-$\ell$ posterior of partially observed CMB maps, both synthetic 
and real.  From this exploration, we built a copula-based
approximation for that posterior distribution.  Numerical evaluation
of that approximation is much faster than the pixel-based computation.
We showed that the approximation is very close to the actual posterior
with an accuracy which is probably sufficient for most cosmological
applications.  For example, on a simple two-parameter pseudo
cosmological model, we found a discrepancy which is negligible with
respect to the width of the posterior mode (figure~\ref{fig:anplots}).

The copula approximation uses two ingredients: a model of marginal
distributions and a correlation matrix.
The marginals are mostly distributed as inverse gammas, as in the
full-sky case, but with different parameters.  
Maybe surprisingly, the correlations between (Gaussianized) multipoles
are found to be quite low ($<10\%)$.  Ignoring them in the toy
cosmological model illustrated by figure~\ref{fig:anplots} does not
change significantly the posterior.
However, when considering the full joint distribution of the
multipoles (as opposed to its \emph{projection} onto the two-parameter toy
model), the correlation is significant: the Kullback divergence from
the true posterior to its copula approximation quadruples if the
correlation is left out.  In both cases however, the Kullback 
divergence remains small.

The main limitation of the proposed approximation is that it requires
an exploration of the posterior to measure the parameters of the
approximation.  We used an adaptive importance sampling algorithm, but
a MCMC algorithm, Gibbs-based
\citep{2004PhRvD..70h3511W} or Hybrid MC-based
  \citep{Taylor:2007p3316} can also be used.  Both methods exhibit
good scaling properties thanks to a smart re-writing of the posterior
and could, if convergence is well controlled, provide estimates at
higher $\ell$.
Indeed, a very recent work, published at the time we were finishing
this paper follows a similar path and demonstrate a Gaussianization
technique based on splines rather than on inverse gamma models 
\citep{Rudjord:2008p4864}.

Another approach would be to determine the parameters of the marginals
directly from the likelihood, without resorting to a sampling-based
exploration.  We are currently working on an analytical derivation of
the approximation which would make it possible to build an
approximation valid for higher $\ell$ at low computational cost.
Being able to reach smaller scales is also important to explore the
transition between low $\ell$ estimates and high-$\ell$ ones.
Indeed, at very small scales, the problem becomes intractable and
requires the use of asymptotic approximations to the
likelihood~\citep{Percival:2006MNRAS.372.1104P,Smith:2006PhRvD..73b3517S}.

Finally, it is not clear yet whether the same kind of approximation
can be built for polarized fields. In the temperature case 
addressed here, we took
advantage of a low correlation situation, thanks to a high signal to
noise ratio and relatively small masked area.  Polarized observations
will be noisier and it remains to be seen if copula approximations are
up to the task.  This is the subject of current investigations.

\section*{Acknowledgments}

We thank J. Dunkley for her detailed description of the large scale
map used in the WMAP5 likelihood.  The authors were greatly helped by
the comments and remarks from F.~Bouchet, H.K.~Eriksen, members of the
ECOSSTAT ANR project and the Planck CTP working group.  The ANR grant
ECOSSTAT (ANR-05-BLAN-0283-04) provided financial support for part of this work.
We acknowledge the use of the HEALPix package\footnote{\href{http://healpix.jpl.nasa.gov}{http://healpix.jpl.nasa.gov}}.

\appendix

\section{ML estimation of inverse gamma parameters}\label{sec:mlegamma}

The log-likelihood $\log\mathcal{L}(\alpha,\beta)$ for a sample of $N$
independent realizations $X_i$ under an inverse gamma density is
\begin{displaymath}
  \log\mathcal{L}= \sum_i^N \left( \alpha\log\beta-\log\Gamma(\alpha)  - (\alpha+1)\log X_i - \frac{\beta}{X_i} \right)
\end{displaymath}   
as seen from eq. (\ref{eq:defigamma}).
The ML estimate for $(\alpha,\beta)$ is the solution of
$\frac{\partial\log\mathcal{L}}{\partial\alpha}=0$ and
$\frac{\partial\log\mathcal{L}}{\partial\beta}=0$ leading to the two
estimating equations:
\begin{displaymath}
  \log\beta-\psi(\alpha)=\frac{1}{N}\sum_i^N\log X_i
  ,
  \qquad
  \frac{\alpha}{\beta}-=\frac{1}{N}\sum_i^N\frac{1}{X_i}
\end{displaymath}
where $\psi(u)$ is the log-derivative of the gamma function, also known
as the digamma function.  Using the last equation to express $\beta$
in terms of $\alpha$, the ML estimate can be obtained by solving
\begin{equation}
  \log\alpha-\psi(\alpha)=\frac{1}{N}\sum_i^N\log X_i - \log \left(\frac{1}{N}\sum_i^N\frac{1}{X_i} \right).
\end{equation}
This is quickly done numerically in a few steps of a Newton
algorithm; both the digamma function and its derivative being
available in the GSL package.

\bibliographystyle{mn2e}
\bibliography{lowly}

\label{lastpage}

\end{document}